\documentclass[twocolumn]{aastex631}  %linenumbers
\usepackage{graphicx}

\shorttitle{GW-MMADS kilonova constraints on S250206dm}
\shortauthors{Hu et al.}

\graphicspath{{./}}
\begin{document}

% \title{Template \aastex Article with Examples: 
% v6.31\footnote{Released on March, 1st, 2021}}
\title{Kilonova constraints for the LIGO/Virgo/KAGRA neutron star merger candidate S250206dm: GW-MMADS observations}

\newcommand{\mcwilliams}{
    McWilliams Center for Cosmology and Astrophysics,
    Department of Physics,
    Carnegie Mellon University,
    5000 Forbes Avenue, Pittsburgh, PA 15213
}

\newcommand{\lmu}{
    University Observatory, 
    Faculty of Physics, 
    Ludwig-Maximilians-Universität München, 
    Scheinerstr. 1, 81679 Munich, Germany
}

\newcommand{\unc}{
    Department of Physics and Astronomy, 
    University of North Carolina at Chapel Hill, 
    Chapel Hill, NC 27599-3255, USA
}

\newcommand{\cbpf}{
    Centro Brasileiro de Pesquisas F\'isicas, Rua Dr. Xavier Sigaud 150, 
    22290-180 Rio de Janeiro, RJ, Brazil  
}

\correspondingauthor{Lei Hu}
\email{leihu@andrew.cmu.edu}

\author[0000-0001-7201-1938]{Lei Hu}
\affiliation{\mcwilliams}
% \email{leihu@andrew.cmu.edu}

\author[0000-0002-1270-7666]{Tom\'as Cabrera}
\affiliation{\mcwilliams}
% \email{tcabrera@andrew.cmu.edu}

\author[0000-0002-6011-0530]{Antonella Palmese}
\affiliation{\mcwilliams}
% \email{palmese@cmu.edu}

\author[0009-0006-7990-0547]{James Freeburn}
\affiliation{Centre for Astrophysics and Supercomputing, Swinburne University of Technology, John St, Hawthorn, VIC 3122, Australia}
\affiliation{ARC Centre of Excellence for Gravitational Wave Discovery (OzGrav), John St, Hawthorn, VIC 3122, Australia}
% \email{jamesfreeburn54@gmail.com}

\author[0000-0002-8255-5127]{Mattia Bulla}
\affiliation{Department of Physics and Earth Science, University of Ferrara, via Saragat 1, I-44122 Ferrara, Italy}
\affiliation{INFN, Sezione di Ferrara, via Saragat 1, I-44122 Ferrara, Italy}
\affiliation{INAF, Osservatorio Astronomico d’Abruzzo, via Mentore Maggini snc, 64100 Teramo, Italy}
% \email{mattia.bulla@unife.it}

\author[0000-0002-8977-1498]{Igor Andreoni}
\affiliation{\unc}
% \email{igor.andreoni@unc.edu}

\author[0000-0002-9364-5419]{Xander J. Hall}
\affiliation{\mcwilliams}
% \email{xjh@andrew.cmu.edu}

\author[0000-0002-9700-0036]{Brendan O'Connor}
\affiliation{\mcwilliams}
% \email{boconno2@andrew.cmu.edu}

\author[0000-0003-3433-2698]{Ariel Amsellem}
\affiliation{\mcwilliams}
% \email{aamselle@andrew.cmu.edu}

\author[0000-0003-4383-2969]{Clécio R. Bom}
\affiliation{\cbpf}
% \email{clecio@debom.com.br}

\author[0009-0001-0574-2332]{Malte Busmann}
\affiliation{\lmu}
% \email{m.busmann@physik.lmu.de}

\author[0009-0006-7670-9843]{Jennifer Fabà}
\affiliation{\lmu}
% \email{J.Faba@campus.lmu.de}

\author{Julius Gassert}
\affiliation{\lmu}
% \email{julius.gassert@campus.lmu.de}

\author{Sena Kalabalik}
\affiliation{\lmu}
% \email{Sena.Kalabalik@campus.lmu.de}

\author[0009-0000-4830-1484]{Keerthi Kunnumkai}
\affiliation{\mcwilliams}
% \email{kkunnumk@andrew.cmu.edu}

\author[0000-0003-3270-7644]{Daniel Gruen}
\affiliation{\lmu}
\affiliation{Excellence Cluster ORIGINS, Boltzmannstr. 2, 85748 Garching, Germany}
% \email{daniel.gruen@lmu.de}

\author[0000-0003-3402-6164]{Luidhy Santana-Silva}
\affiliation{\cbpf}
% \email{luidhysantana@gmail.com}

\author[0000-0002-1420-3584]{André Santos}
\affiliation{\cbpf}
% \email{andsantos@cbpf.br}

% alphabetical after this line

\author[0000-0002-2184-6430]{Tomás Ahumada}
\affiliation{Division of Physics, Mathematics and Astronomy, California Institute of Technology, Pasadena, CA 91125, USA}
% \email{tahumada@astro.caltech.edu}

\author[0000-0001-8544-584X]{Jonathan Carney}
\affiliation{\unc}
% \email{jcarney@unc.edu}

\author[0000-0002-8262-2924]{Michael W. \surname{Coughlin}}
\affiliation{School of Physics and Astronomy, University of Minnesota, Minneapolis, MN 55414}
% \email{cough052@umn.edu}

\author[0000-0003-3021-4897]{Xingzhuo Chen}
\affiliation{George P. and Cynthia Woods Mitchell Institute for Fundamental Physics \& Astronomy, \\
Texas A. \& M. University, Department of Physics and Astronomy, 4242 TAMU, College Station, TX 77843, USA}
\affiliation{Texas A\&M Institute of Data Science
John R. Blocker Building, Suite 227
155 Ireland Street, TAMU 3156
College Station, TX 77843-3156}
% \email{chenxingzhuo@tamu.edu}

\author[0000-0002-5956-851X]{K. E. Saavik Ford}
\affiliation{Center for Computational Astrophysics, Flatiron Institute, 
162 5th Ave, New York, NY 10010, USA}
\affiliation{Department of Astrophysics, American Museum of Natural History, New York, NY 10024, USA}
\affiliation{Department of Science, BMCC, City University of New York, New York, NY 10007, USA}
% \email{sford@amnh.org}

\author[0000-0002-0175-5064]{Daniel E. Holz}
\affiliation{Kavli Institute for Cosmological Physics, University of Chicago, Chicago, IL 60637, USA} 
\affiliation{Enrico Fermi Institute, University of Chicago, Chicago, IL 60637, USA}
\affiliation{Department of Physics, University of Chicago, Chicago, IL 60637, USA}
\affiliation{Department of Astronomy and Astrophysics, University of Chicago, Chicago, IL 60637, USA}
% \email{holz@uchicago.edu}

\author[0000-0002-5619-4938]{Mansi M. Kasliwal}
\affiliation{Division of Physics, Mathematics, and Astronomy, California Institute of Technology, Pasadena, CA 91125, USA}
% \email{mansi@astro.caltech.edu}

\author[0000-0003-2362-0459]{Ignacio Maga\~na~Hernandez}
\affiliation{\mcwilliams}
% \email{imhernan@andrew.cmu.edu}

\author{Cassidy Mihalenko}
\affiliation{School of Natural Sciences, University of Tasmania, Private Bag 37 Hobart, Tasmania, 7001, Australia}
\affiliation{ARC Centre of Excellence for Gravitational Wave Discovery (OzGrav), John St, Hawthorn, VIC 3122, Australia}
% \email{cassidy.mihalenko@utas.edu.au}

\author[0000-0002-3635-5677]{Rosalba Perna}
\affiliation{Department of Physics and Astronomy, Stony Brook University, Stony Brook, NY 11794-3800, USA}
% \email{rosalba.perna@stonybrook.edu}

\author[0000-0002-5466-3892]{Arno Riffeser}
\affiliation{\lmu}
\affiliation{Max-Planck-Institut für Extraterrestrische Physik, Giessenbachstraße 1, 85748 Garching, Germany}
% \email{arri@usm.lmu.de}

\author{Christoph Ries}
\affiliation{\lmu}
% \email{cries@usm.lmu.de}

\author{Lena Schnappinger}
\affiliation{\lmu}
% \email{l.schnappinger@campus.lmu.de}

\author{Michael Schmidt}
\affiliation{\lmu}
% \email{mschmidt@usm.lmu.de}

\author[0000-0002-1154-8317]{Julian Sommer}
\affiliation{\lmu}
% \email{Julian.Sommer@campus.lmu.de}

\author{Sarah Teague}
\affiliation{\unc}
% \email{steague1@unc.edu}

\author{Pablo Vega}
\affiliation{\lmu}
% \email{P.Vega@campus.lmu.de}

\author[0009-0008-1415-6678]{Olga Volchansky}
\affiliation{Department of Orthopaedics, 
    University of North Carolina at Chapel Hill, 
    Chapel Hill, NC 27599-3255, USA}
% \email{olga_volchansky@med.unc.edu}

\author[0000-0001-7092-9374]{Lifan Wang}
\affiliation{George P. and Cynthia Woods Mitchell Institute for Fundamental Physics \& Astronomy, \\
Texas A. \& M. University, Department of Physics and Astronomy, 4242 TAMU, College Station, TX 77843, USA}
% \email{lifan@tamu.edu}

\author[0000-0003-2976-8198]{Yajie Zhang}
\affiliation{\lmu}
% \email{yajie.zhang@campus.lmu.de}



\begin{abstract}

Gravitational wave (GW) neutron star mergers with an associated electromagnetic counterpart constitute powerful probes of binary evolution, the production sites of heavy elements, general relativity, and the expansion of the universe. Only a handful of candidate GW binary mergers during the fourth LIGO/Virgo/KAGRA observing run (O4) so far are believed to include a neutron star. We present optical-near infrared follow-up observations of the candidate neutron-star black hole GW merger S250206dm. This is the first high-significance mass gap neutron star-black hole candidate observed by multiple GW detectors (thus having a significantly smaller sky localization than one-detector events), offering the first opportunity to effectively follow up a GW event of this kind. Our GW MultiMessenger Astronomy DECam Survey (GW-MMADS) campaign consisted of a wide-field search using the Dark Energy Camera (DECam) and T80-South (T80S), as well as galaxy-targeted observations using the Southern Astrophysical Research (SOAR) imager and the Wendelstein 2.1m 3-channel camera. No viable kilonova counterpart was found in our observations. 
We use our observation depths to place competitive constraints on kilonova models similar to or brighter than the GW170817 kilonova AT 2017gfo within our observed fields, ruling out 100\% of such models with SOAR galaxy-targeted observations and $\sim43$\% (48\%) with DECam (DECam and T80S). 

\end{abstract}

% \keywords{Classical Novae (251) --- Ultraviolet astronomy(1736) --- History of astronomy(1868) --- Interdisciplinary astronomy(804)}


\section{Introduction} 

Multimessenger observations of gravitational wave (GW) events can enable a wide range of scientific analyses, from stellar evolution studies to measurements of fundamental physical and cosmological parameters. This has been exemplarily demonstrated with the first multimessenger GW detection, GW170817 \citep{ligobns}, observed by the Laser Interferometer Gravitational-wave Observatory (LIGO; \citealt{2015LIGO}), Virgo \citep{Acernese_2014}, and a large number of multiwavelength EM facilities \citep{MMApaper}. GW events arising from binary neutron star (BNS) or black hole-neutron star (BHNS) mergers can give rise to a variety of electromagnetic (EM) counterparts at different wavelengths. Notably, in the optical to infrared range, these objects can emit EM radiation from a kilonova (KN; e.g. \citealt{metzger_kilonovae}), a transient powered by the radioactive decay of heavy $r$-process elements produced in the merger ejecta \citep{Lattimer1974}. At the time of writing, one high-confidence BNS GW merger was detected after GW170817 \citep{GW190425Abbott2020}, but no EM counterpart was found for this event, likely due to its larger distance and sky localization \citep{GW190425Coughlin2019}. % other options for citation: GW190425Hosseinzadeh2019, GW190425Coulter2024

Depending on the compact objects' masses, the neutron star equation of state, and the black hole spin, BHNS mergers can also give rise to a kilonova \citep{PhysRevD.86.124007,PhysRevD.88.041503,Fernandez2017,Kawaguchi2016,Kasen2015}. BHNS binaries with a low mass black hole are more likely to produce a KN, while higher mass black holes cause the NS to fall into the black hole without significant disruption and therefore little to no EM emission. Because black hole observations predating those by LIGO and Gaia, predominantly based on EM data, indicate that black holes typically have masses $>5~M_\odot$, a ``mass-gap'' between the neutron star and the black hole population was conjectured to exist between $3\mbox{--}5~M_\odot$. A dearth of BHNS mergers with black holes in that mass range would render multimessenger prospects for BHNS extremely challenging. Follow-up campaigns for BHNS with black hole mass $>5~M_\odot$ have not yielded any EM counterpart discovery \citep{anand2021}. The fourth LIGO/Virgo/KAGRA observing run (O4; 2023-2025) recently revealed the first BHNS likely containing a mass-gap black hole, GW230529 \citep{230529_LVK}. The existence of mass-gap BHNS mergers renders BHNS mergers an invaluable source of multimessenger kilonova detections \citep{kunnumkai_230529,kunnumkai_O4O5}. Unfortunately only one GW detector was online at the time of GW230529, so that its sky localization is extremely poor, challenging most EM follow up efforts. The non-detection of a Gamma-Ray Burst enabled constraints on the jet luminosity and opening angle \citep{Ronchini_2024}, while some kilonova constraints were only possible for a small fraction of the sky and models \citep{2025arXiv250315422P}.

On February 6, 2025 at 21:25:30 UT, the LIGO and Virgo detectors observed S250206dm \citep{2025GCNLVK,2025GCNLVK2,2025GCNLVK1,2025GCNLVK3}, a candidate GW event with a marginalized luminosity distance of 373 $\pm$ 104 Mpc and a sky localization of 38 and 547 $\mbox{deg}^2$ at 50 and 90\% credible interval respectively. With a 55\% probability of being a BHNS merger, a 37$\%$ chance of being a BNS (based on the latest \texttt{pycbc} alert estimates), as well as a 62\% probability of hosting a black hole in the mass gap and a 30\% chance of having disrupted material outside of the innermost stable circular orbit of the black hole, this is likely the most promising multimessenger source in O4 thus far. Other GW alerts during O4 have not been confidently associated with any EM counterpart \citep{cabrera24,Ahumada:2024qpr,2025arXiv250315422P,2025arXiv250602224D}. No gamma-ray emission associated with S250206dm was detected by Fermi-GBM \citep{S250206dm_NoGRB}.

In this work, we present the follow-up campaign carried out as part of the Gravitational Wave MultiMessenger DECam Survey (GW-MMADS; Proposal ID 2023B-851374, PI: Andreoni \& Palmese), in coordination with other facilities, namely the Southern Astrophysical Research (SOAR) 4m, the T80S 0.8m, and the Wendelstein 2.1m telescopes, and the South African Large Telescope (SALT). A companion work (Ahumada et al., in prep.) described the campaign and results from the Zwicky Transients Facility (ZTF), which, combined with ours, covers the majority of the high probability region for this event. See also \citet{frostig2025winter} for near-infrared constraints on this event. In \S\ref{sec:obs} we describe our follow-up strategy and observations, in \S\ref{sec:candidate} we present the candidates identified from the campaign, in \S\ref{sec:model} we constrain kilonova models based on our findings, and in \S\ref{sec:conclusion} we report our conclusions.

\section{Observations} \label{sec:obs}

\begin{figure*}[ht!]
    \includegraphics[trim=1.5cm 0cm 2cm 0cm,clip=true,width=17cm]{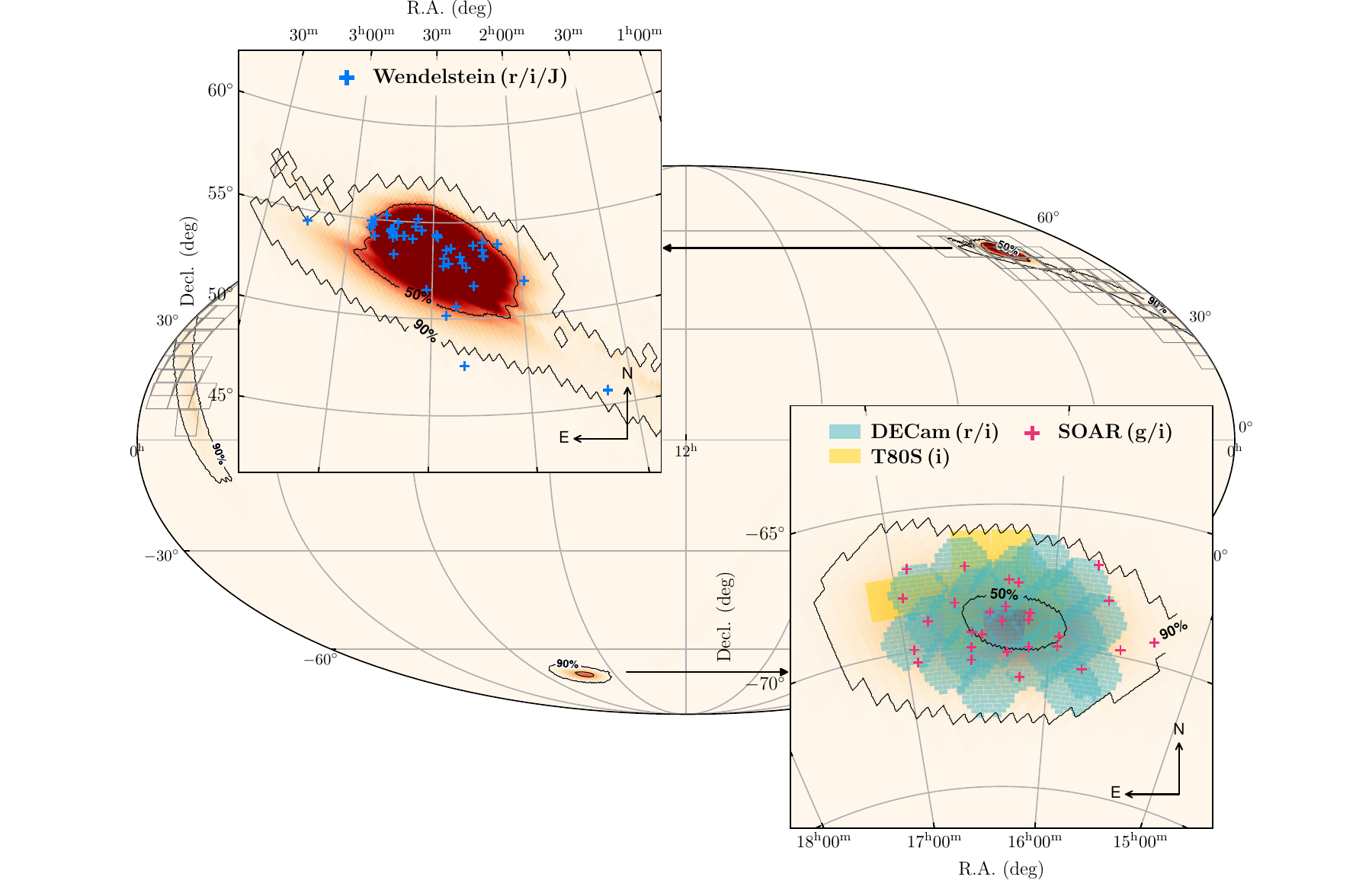}
    \caption{Localization region of S250206dm overlapped with the footprints of GW-MMADS follow-up observations. The upper and lower inset panels show zoomed-in views of two lobes of the LVK skymap with high probability, located in the Northern and Southern Hemispheres, respectively. For reference, we also overlay in gray the sky tiles in the northern lobe used for ZTF observations of S250206dm (Ahumada et al., in prep.).}\label{fig:skymap}
\end{figure*}

\subsection{Galaxy-targeted Search with SOAR/Goodman} \label{subsec:soar}

Approximately 9 hours after the GW trigger, we conducted a galaxy-targeted search using the Goodman High Throughput Spectrograph (HTS; \citealt{Clemens2004}) mounted on the Southern Astrophysical Research (SOAR) 4.1 m telescope\footnote{\url{https://noirlab.edu/science/programs/ctio/telescopes/soar-telescope}}, operated in imaging mode for Target of Opportunity (ToO) observations (PI: Andreoni).
% \textcolor{blue}{For Jim/Igor: The following are mainly summarized from SOAR's two GCNs, you may want to refine the descriptions.}
We observed 33 galaxies located in the highest-probability region of S250206dm in both $g$ and $i$ bands \citep{SOAR_GCN1}, with corresponding pointings shown in Figure~\ref{fig:skymap}.
These galaxies were selected by cross-matching the sky localization with the NED Local Volume Sample \citep[NED-LVS;][]{NED_LVS_Cook2023}, taking into account the NED galaxy list of S250206dm from \citet{NED6_GCN}.
Data acquisition started on 2025 February 7 at 06:36:12.51 UT and lasted for about two hours, reaching typical depths of 22.2 and 21.7 mag for $g$ and $i$ bands, respectively. 
On 2025 February 9, SOAR revisited 30 galaxies in the highest probability region, conducting $i$-band observations from 06:48 UT to 09:18 UT with comparable depths as the previous night \citep{SOAR_GCN2}. The details of SOAR pointings are available in TreasureMap\footnote{\url{https://treasuremap.space/alerts?graceids=S250206dm}}.

\subsection{Galaxy-targeted Search with Wendelstein} \label{subsec:wendelstein}

In the Northern Hemisphere, a galaxy-targeted search for the optical and near-infrared counterpart of S250206dm was initialized with the Three Channel Imager (3KK; \citealt{2016SPIE.9908E..44L}) on the 2.1 m Fraunhofer Telescope at Wendelstein Observatory (FTW; \citealt{2014SPIE.9145E..2DH}). %located on Mt.~Wendelstein at the northern edge of the Alps. 
The 3KK imager enables simultaneous observations in three channels with a $7^\prime \times 7^\prime$ FoV. For S250206dm, we configured the $r$, $i$ and $J$ bands for blue, red and near-infrared channels, respectively. 
Beginning 19 hr after the event, on 2025 February 7 at 19:00:21 UT, Wendelstein 3KK commenced observations on 50 targets selected from the NED Galaxies in the Localization Volume in \cite{NED6_GCN}. The selection was based on ranking galaxies by the product of the 3D localization probability and WISE W1 luminosity, retaining only targets located in the Northern Hemisphere (with Decl. $>$ 0).
On the same night, Wendelstein 3KK in addition observed the 90$\%$ localization region of EP-WXT trigger 01709131361, which lies within the 70$\%$ localization of S250206dm, beginning at 2025 February 7 at 17:52:05 UT (20.4 hours after the merger) \citep{GCN_Wendelstein_EP}.
% \textcolor{red}{For Jennifer/Sena: According to the table of limiting magnitude, only 86 exposures in the first night - seems not aligned with 50 galaxies in three bands?} 
On 2025 February 8, Wendelstein observations proceeded with observations of 12 targets from the updated NED Galaxies in the Localization Volume \citep{NED7_GCN}, including 11 from the original sample and one new addition. We also followed up the candidate AT~2025baz reported by ZTF \citep{GCN_2025baz}.
In the subsequent weeks, intermittent follow-ups were conducted on a subset of transient candidates. 
Figure~\ref{fig:skymap} shows the footprints of the galaxy-targeted search with the Wendelstein 3KK observations overlaid on the LVK skymap of S250206dm. The details of Wendelstein pointings are available in TreasureMap\footnote{Observations affected by poor weather and resulting in unsuccessful data reduction were excluded from the TreasureMap pointings.}. 

\subsection{DECam Search}

Since S250206dm occurred on the day the DECam camera was taken off the Blanco 4-meter telescope at Cerro Tololo Interamerican Observatory (CTIO) for engineering, DECam was unavailable during the early follow-up phase of the event. GW-MMADS commenced ToO observations with DECam immediately following the camera's reinstallation. The DECam observations began on 2025 February 13 at 06:15:00.30 UT -- approximately six days after the binary merger -- and lasted for three hours.

On the first night, we adopted an exposure sequence of $i$-$r$-$i$ band for each pointing, with exposure times of 60\,s in the $r$ band and 140\,s in the $i$ band, reaching typical depths of 22.6 and 23.0 mag, respectively. 
The relatively deep $i$-band observations were designed to enhance the detection sensitivity to a potential red counterpart associated with the GW event at this phase. 
To minimize contamination from moving objects, we maintained a typical gap of 90 minutes between two $i$-band observations of each pointing.
As shown in Figure~\ref{fig:skymap}, we used 23 pointings to fully cover the 50\% probability region and the majority of the 90\% probability region in the southern hemisphere. 
We note that our DECam coverage is slightly redundant, particularly within the 50\% probability region, as the pointings were intentionally selected with significant overlap to mitigate the risk of candidate loss due to CCD chip gaps.
The details of DECam pointings are available in TreasureMap.

We repeated the observations on the following night using the same strategy, with a slightly longer $r$-band exposure of 70\,s, beginning on 2025 February 14 at 06:22:17.23 UT.
DECam follow-up observations resumed three days later, starting on 2025 February 17 at 07:48:50.60 UT, consisting of a single round of $i$-band imaging with the same 140\,s exposure time.
The final DECam observations were conducted on 2025 February 27, beginning at 07:04:25.54 UT, using the same $i$-$r$-$i$ exposure sequence as the first night, but with increased exposure times of 280\,s in the $i$ band and 70 s in the $r$ band.

\subsection{T80-South search}

We conducted observations using the T80Cam on board the T80-South Telescope located at CTIO, triggered via a Target of Opportunity (ToO) approximately 9 hours after the gravitational wave alert. We observed 9 tiles inside the lobe located in the southern hemisphere, using 300-second exposures in the $i$-band. This observation strategy provided a total coverage of 18 deg$^2$ during the night. The choice of exposure time and filter was optimized to enhance sensitivity to a potential kilonova emission expected in the early times post-merger.
The tiling of the GW skymap was performed using {\tt Teglon} \citep{tegloncoulter_2021}, an open-source tool developed for optimized planning of electromagnetic follow-up observations of GW events. {\tt Teglon} improves upon the original LVK localization by incorporating a completeness metric derived using a galaxy catalog - in our case, the GLADE catalog \citep{Dalya2018} - to estimate the spatial distribution of potential host galaxies in three dimensions. This results in a reweighted localization map that accounts for galaxy density and distance priors.
{\tt Teglon} then subdivides the updated probability map into fields corresponding to the field of view of the T80-South telescope, generating a ranked list of pointings for observations. However, after our observing sequence was submitted to the T80-South queue, an updated version of the GW skymap was released. This revised localization slightly shifted the position of the highest-probability region within the southern lobe. As a result, a fraction of the most probable sky area was not included in our executed pointings.

\section{Candidates} \label{sec:candidate}

\subsection{Candidates from SOAR/Goodman}

We performed image differencing analysis for the SOAR/Goodman observations using Saccadic Fast Fourier Transform \citep[SFFT\footnote{\url{https://github.com/thomasvrussell/sfft}};][]{SFFT2022}. Two approaches were implemented to search for a potential counterpart on the difference images: (i) using the SOAR/Goodman observations from the first epoch (2025 February 7) as science images and the observations from second epoch (2025 February 9) as templates \citep{SOAR_GCN2}; (ii) subtracting SOAR images of both epochs using the best available archival DECam images as templates. 
All SOAR/Goodman images were photometrically calibrated to SkyMapper DR4 \citep{SkyMapper_2024}. For the first approach, candidates were identified by visual inspection of each difference image. A total of 22 candidates were initially flagged. After removing artifacts caused by bad subtractions, cosmic rays and known variable stars, only one candidate remained: AT~2025ber. Near-infrared follow-up observations of AT~2025ber was carried out by NEWFIRM on 2025 February 12, resulting in a non-detection at the site of the candidate, with a 3$\sigma$ depth of $H\sim23$ AB magnitude \citep{2025GCN.39303....1C}. 
We therefore assessed AT2025ber to be spurious or a moving object, likely not associated with S250206dm. Visual inspection was also performed for the second approach on the difference images from two epochs of SOAR observations. No convincing transient-like signals were identified. 

\subsection{Candidates from Wendelstein}

Wendelstein data were reduced using a custom analysis pipeline based on \citet{2002A&A...381.1095G}, which applies standard image processing, including the common CCD-level corrections. The calibrated images were subsequently stacked using the AstrOmatic software suite \citep{SExtractor, Bertin2002, Bertin2006}. 
To search for variable sources, image differencing analysis was conducted on the stacked images using SFFT, employing Wendelstein images taken from the same field as reference, mostly at a late phase in March. 
No variability was detected upon careful inspections, except for a nuclear transient candidate (with internal name $\textrm{S250206dm\_022200\_p503737}$) at R.A. = 02:22:00.28, decl. = +50:37:37.10.
However, a brightening of the transient candidate was detected on March 8, which does not match the expected photometric behavior of a kilonova, suggesting it may be related to AGN variability.
For the EP-WXT trigger 01709131361, we did not identify any objects that brightened with respect to archival Legacy Survey $r$-band data or 2MASS $J$-band data, except for the flaring star reported by \cite{2025GCN.39218....1L}, which is likely the progenitor of the EP-WXT trigger.

\subsection{Candidates from DECam}

We analyzed the DECam observations using our GPU-enabled image differencing pipeline for rapid transient detection and photometry (Hu et al., in prep.), based on the high-efficiency SFFT algorithm \citep{SFFT2022}. The DECam data processing and candidate vetting for S250206dm using our pipeline is summarized as follows:
\begin{itemize}
    \item For each scheduled DECam pointing in a specific band, we retrieved the best available archival DECam images with overlapping coverage from the NOIRLab data archive. Multiple exposures were requested to enable subsequent image stacking, ensuring sufficient depth for transient detection in our data. For the case of S250206dm, we typically retained a reference coverage contributed from five archival DECam exposures. Reference catalogs from Gaia DR3 \citep{GAIA_DR3_2022} and DELVE \citep{DELVE_2021} were also obtained for photometric calibration and source classification. These preparatory steps were carried out once the observing plan was finalized.
    
    \item The pipeline acquired newly observed DECam images calibrated by the NOIRLab DECam Community Pipeline (CP; \citealt{2014ASPC..485..379V}) during the night of observations. Each image was photometrically calibrated to Gaia DR3 with a color transformation from Gaia to DECam following a polynomial-fitting prescription introduced in \cite{2020ASPC..527..701G}. 
    % \footnote{Our pipeline prioritizes photometric calibration using Legacy Survey \citep{2019AJ....157..168D}, followed by Pan-STARRS \citep{2016arXiv161205560C} and Gaia. As neither of the first two were available in this case, Gaia DR3 was used for photometric calibration.}
    
    \item Rapid image differencing was then performed using SFFT on each science image and its overlapping individual reference images, after alignment with SWarp \citep{SWarp}. The resulting difference images for each pointing and filter were median-combined per night to generate a deep difference image, improving sensitivity of transient detection\footnote{This approach is robust, albeit less computationally efficient than performing image differencing between stacked reference and science images. Since the reference exposures are generally not aligned with one another, stacking them can introduce significant PSF discontinuities—particularly across chip gaps—which are typically not directly handled by image differencing algorithms.}. We note that the difference images were normalized to a common photometric zero-point before stacking.

    \item We preformed transient detection on the stacked difference images using SourceExtractor \citep{SExtractor} to generate an initial pool of candidate detections. We first removed the detections contaminated by bad pixels (e.g., due to saturated sources) recorded in the DECam CP data quality mask products. To further eliminate artifacts, we employed a rotation-invariant convolutional neural network-based real/bogus classifier, adapted from the implementation in \citet{2022FrASS...9.7100S} and retrained on DECam data. Detections with a real/bogus score $\geqslant 0.3$ are considered transient alerts by our pipeline. 
    
    \item Our pipeline executed an automated filtering process on transient candidates with at least one alert. 
    (i) {\bf Not stellar}: We excluded stellar objects by cross-matching with the Gaia DR3 and DELVE catalogs, removing sources flagged as stars based on proper motion in Gaia and morphological classification in DELVE. 
    (ii) {\bf Not MPC}: The pipeline accessed the Minor Planet Center (MPC) database to eliminate known moving objects. 
    (iii) {\bf Not artifact}: We then required that each transient candidate include at least one high-confidence detection with a real/bogus score $\geqslant 0.7$. 
    The results of the automated filtering process are outlined in Table~\ref{tab:vetting}. We note the automated vetting process eliminated over 90$\%$ of the initial transient candidates.
    
    \item We performed forced aperture photometry on the stacked difference images for each candidate that passed the automated vetting to extract its light curves. The centroid of the detection with the highest signal-to-noise ratio (SNR) was adopted as the center for forced photometry with a fixed aperture of six pixels.
    
    \item Finally, we applied a series of event-based vetting criteria to the remaining candidates to identify potential counterparts of S250206dm. These criteria included: 
    (i) {\bf Early detection}: The candidate must have been discovered during the first night of observations.\footnote{Our DECam observations reached similar depths across the first three epochs; thus, requiring detection on the first night is a reasonable criterion for identifying a fading counterpart.}
    (ii) {\bf Multiple detections}: Candidates with only one detection in both the stacked and individual difference images were excluded, as such detections are likely spurious or caused by moving objects.
    (iii) {\bf Visual inspection}: We visually inspected the subtractions and light curves to further exclude artifacts and probable variable stars. 
    (iv) {\bf Fading behavior}: We computed the steepest decline rate in the candidate's light curve across the four DECam epochs and retained only those with a decline rate $>$ 0.3 mag/day.
    (v) {\bf No prior variability}: We checked the DECam archive, ATLAS forced photometry, and the Transient Name Server (TNS) to ensure there was no previous variability for the candidate.
    (vi) {\bf Red color}: We required a red color, $r - i > 0$, in the light curve after correcting for Milky Way extinction, as expected for a plausible kilonova at 6 days post merger.
    The results of the event-based filtering are summarized in Table~\ref{tab:vetting}.
\end{itemize}
No compelling candidates meeting all vetting criteria were identified in the DECam observations for S250206dm. The 37 candidates that passed visual inspection are listed in the Appendix~\ref{Appendix:DECam_Candidates}, with several were circulated via GCN \citet{2025GCN_GWMMADS}. SALT spectroscopic follow-up observations were conducted for four of these DECam candidates (see Appendix~\ref{Appendix:SALT_Followup} for details).

\subsection{Candidates from T80-South}

All exposures acquired with T80-South were processed using the S-PLUS Transient Extension Program pipeline \citep{Santos2024}. The raw images from the T80-South telescope went through a pre-reduction process that includes bias subtraction, flat-field correction, overscan removal, and image trimming. Once pre-processed, all images go through the reduction pipeline, which applies a nonlinear astrometric correction over the images and performs zero-point calibrations. In order to find variable sources, we perform difference imaging using {\tt hotpants} \citep{Becker2015} and DECam archival images as templates. After image subtraction and inspection of detailed candidate vetting, no transient sources consistent with an EM counterpart were identified within our search region. Given our criteria for finding viable candidates for the following epoch, observations on the following night were not acquired.

\begin{deluxetable}{ccc}
    \tablecaption{
        Candidate filtering results for DECam search. 
        \label{tab:vetting}
    }
    \tablehead{ & Number & Fraction \\
    Vetting Filter & passed & passed}
    \startdata
        \multicolumn{3}{c}{\textit{Regular pipeline cuts}} \\
        \hline
        Alert produced on difference image & 608,054 & 1.00 \\
        Not stellar in Gaia/DELVE & 77,094 & 0.115 \\
        Not MPC object & 77,094 & 0.115 \\
        Real/bogus score $\geqslant$ 0.7 & 40,645 & 0.061 \\ %$\ge 1$ detection with Real/bogus score $>$ 0.9
        \hline 
        \multicolumn{3}{c}{\textit{Event-based cuts}} \\
        \hline
        Discovered in the first night & 16,750 & 0.025 \\
        Multiple detections & 3,589 & 0.005 \\
        Visual inspection & 37 & $6.1 \times 10^{-5}$ \\
        Fast decline in the light curve & 2 & $3.0 \times 10^{-6}$ \\
        No previous history & 1 & $1.5 \times 10^{-6}$ \\
        Red color in the light curve & {\bf 0} & 0 \\
    \enddata
    \tablecomments{The quantities listed in the table represent the number of objects that passed each step in the filtering process. 
    Due to limited coverage of the observed sky fields, the MPC database was insufficient for removing moving objects in this case.}
\end{deluxetable}

\begin{figure*}[ht!]
\centering
    \includegraphics[trim=0.5cm 0cm 0cm 0cm,clip=true,width=18cm]{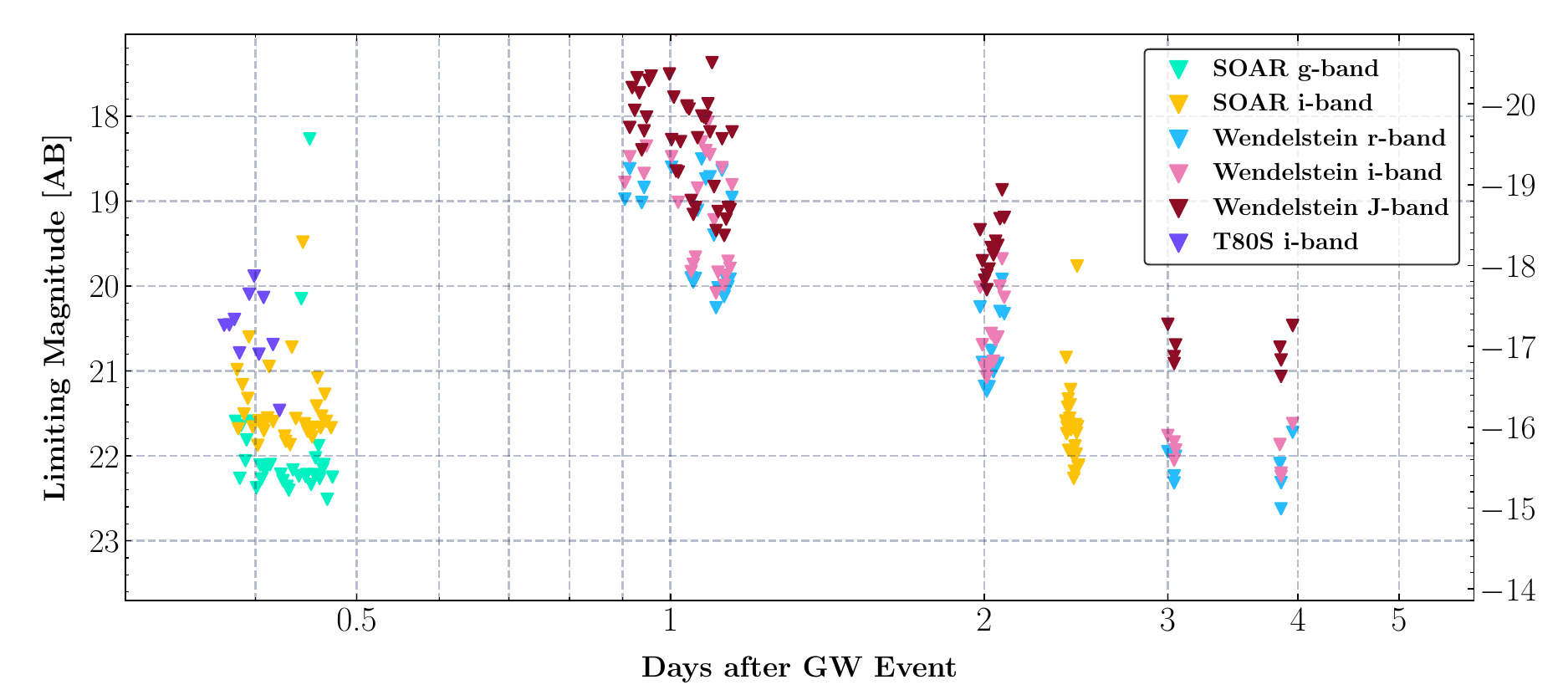}
    \includegraphics[trim=0cm 0cm 0cm 0cm,clip=true,width=18.5cm]{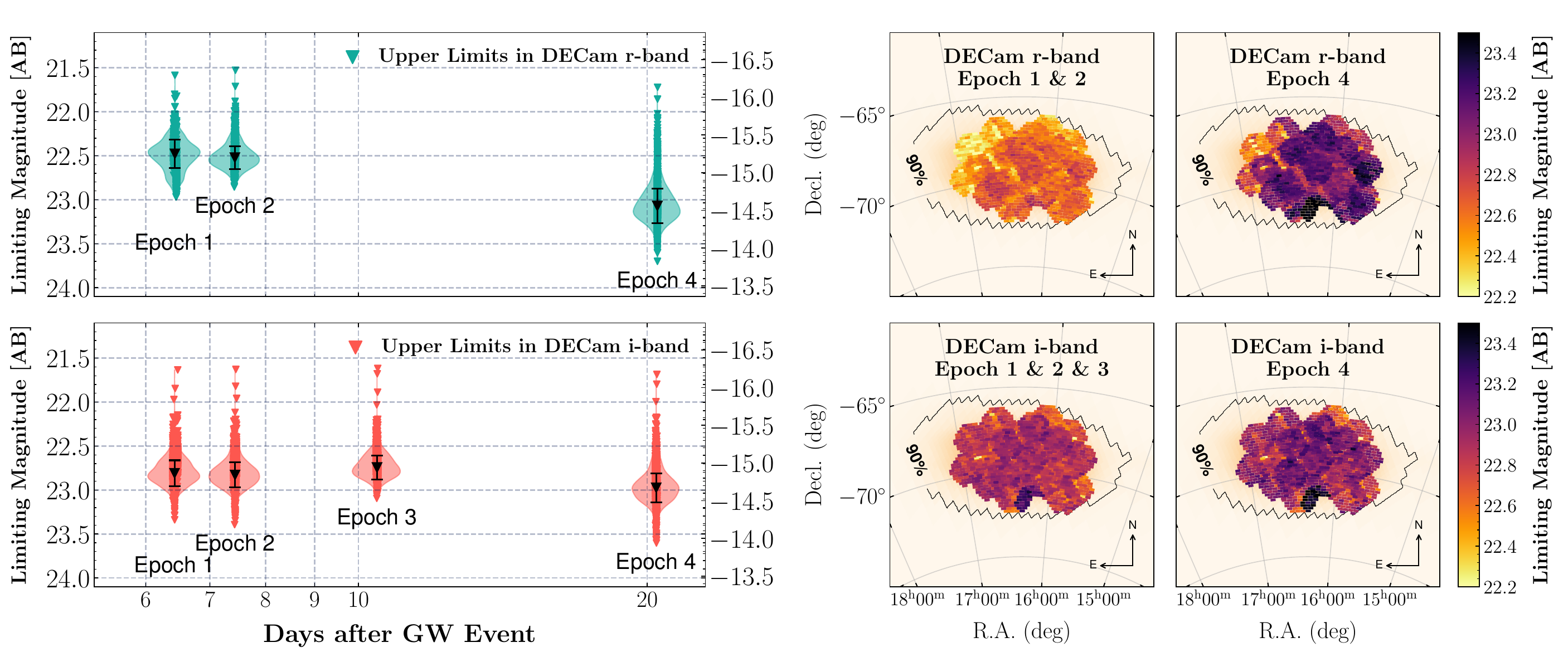}
    \caption{Limiting magnitudes of GW-MMADS observations for S250206dm. \emph{Upper panel:} Limiting magnitudes of early ($<5$ days from trigger) observations as a function of time for S250206dm. Colored triangles represent upper limits obtained with different instruments and filters, including SOAR in $g$/$i$ bands, T80S in $i$-band, and Wendelstein 3KK in $r$/$i$/$J$ bands. For reference, the right y-axis converts apparent magnitudes to absolute magnitudes assuming that S250206dm is at 373 Mpc.
    \emph{Lower left panels:} DECam limiting magnitudes as a function of time after the GW event for DECam $r$-band (top) and $i$-band (bottom). For each epoch, a violin plot is overlaid to represent the distribution of limiting magnitudes across the observations, while a black triangle marks the median upper limit, with error bars indicating the standard deviation. Again, the right-hand y-axis presents absolute magnitudes for a distance of 373 Mpc.
    \emph{Lower right panels:} Spatial variation of limiting magnitudes across the DECam coverage for $r$-band (top) and $i$-band (bottom). For visualization purposes, we grouped epochs with similar depth distributions (e.g., epochs 1 and 2 in the $r$-band) and presented them in a single panel. \label{fig:limmag}}
\end{figure*}

\begin{figure*}[ht!]
    \includegraphics[trim=0.5cm 0cm 0cm 0cm,clip=true,width=17.5cm]{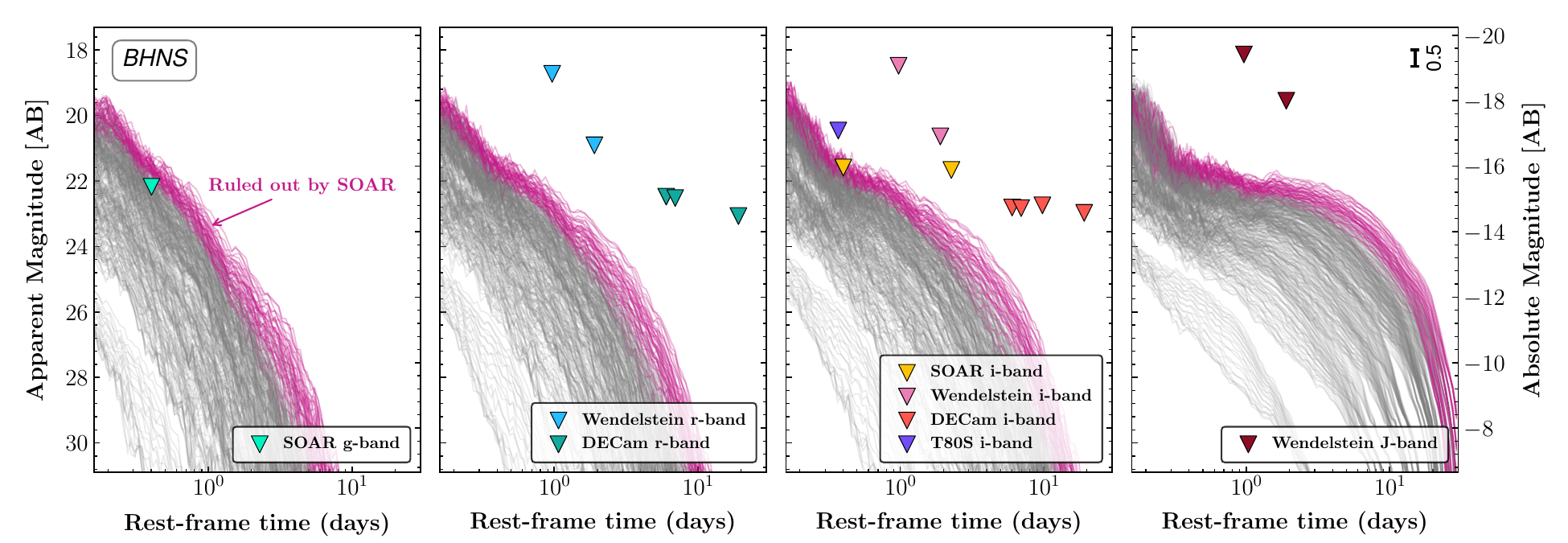}
    \includegraphics[trim=0.5cm 0cm 0cm 0cm,clip=true,width=17.5cm]{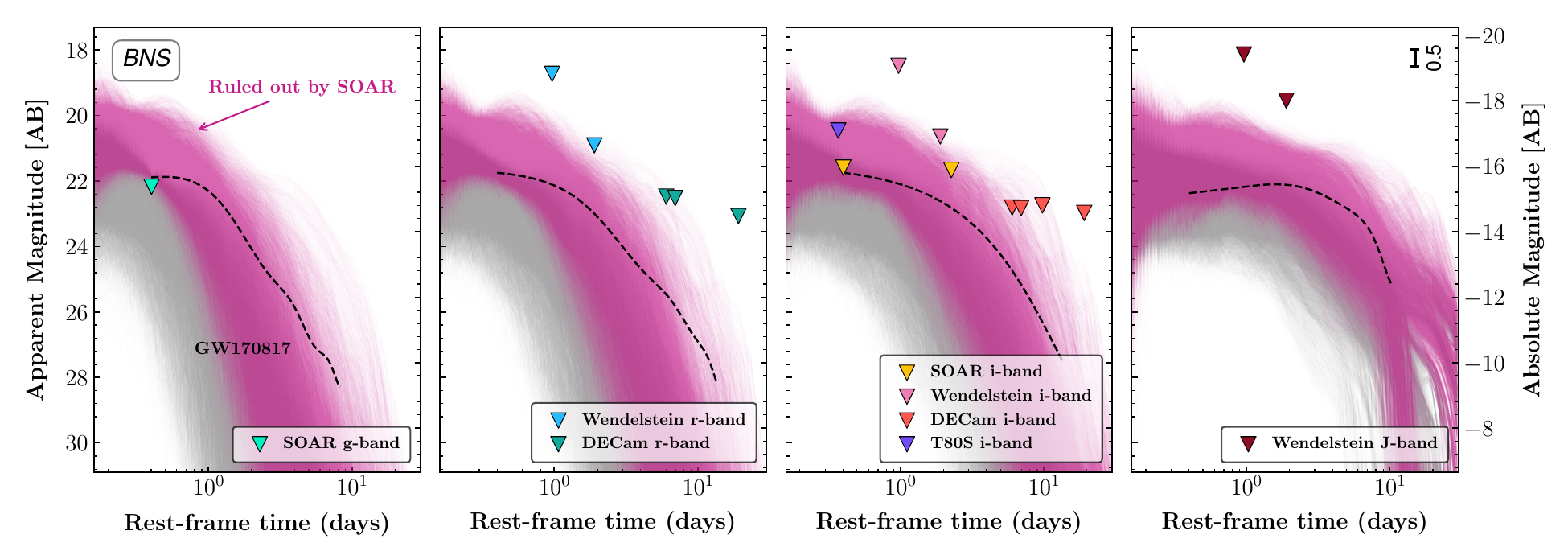}
    \includegraphics[trim=0.5cm 0cm 0cm 0cm,clip=true,width=17.5cm]{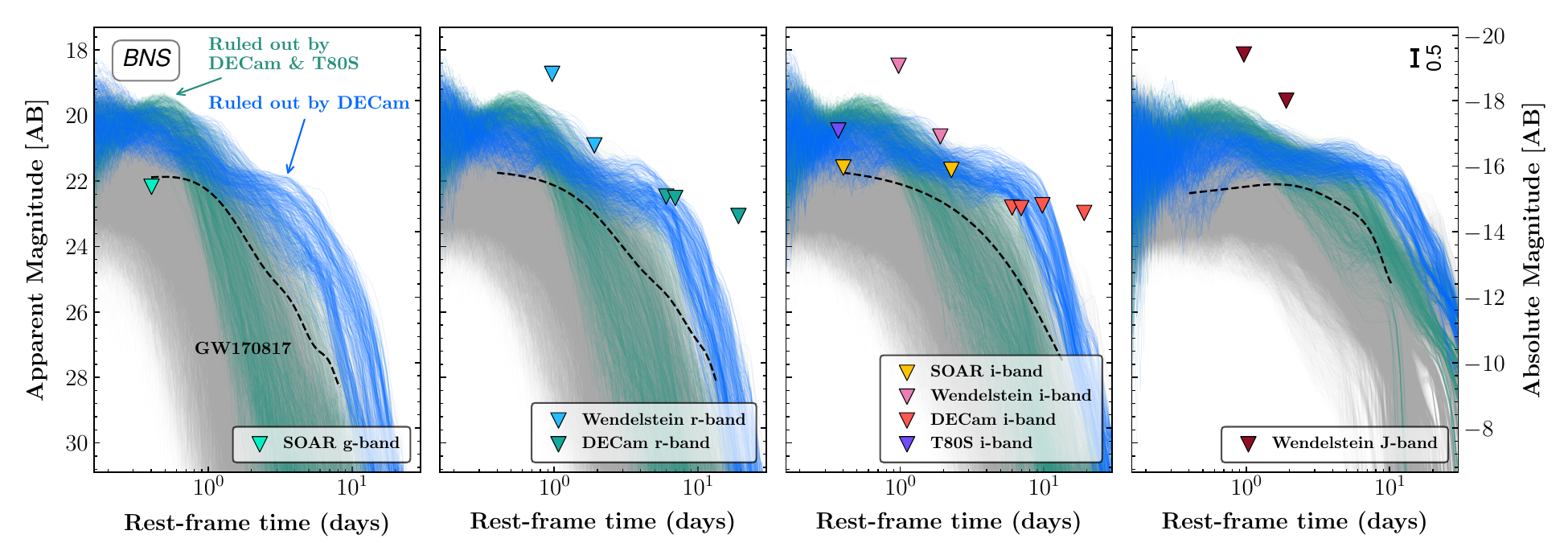}
    \caption{Kilonova light curve simulations for BHNS (top row) and BNS (middle and bottom rows) at a distance of 323 Mpc.
    Each panel shows rest-frame synthetic light curves in a different photometric band, with observed upper limits overlaid as colored triangles. 
    The full ensemble of model predictions is shown in gray. Magenta-highlighted curves indicate models ruled out by SOAR/Goodman galaxy-targeted observations, while blue-highlighted curves mark models disfavored by DECam observations. Green-highlighted curves correspond to the models additionally ruled out jointly by the addition of T80S observations to the DECam ones.
    For reference, the light curves of AT 2017gfo (scaled to a distance of 323 Mpc) are shown as dashed black lines. 
    The multi-band photometry data of AT 2017gfo are taken from \citet{2017ApJ...851L..21V}, and a simple Gaussian process fit is applied to produce smoothed curves for display purposes. 
    We note that uncertainties in the distance translate into vertical shifts in the apparent magnitudes of the model light curves. To illustrate this, we include a 0.5 mag reference scale in the corner of the last panel of each row, representing to first order the apparent magnitude shift corresponding to a 1$\sigma$ uncertainty in distance.}
    \label{fig:lightcurves}
\end{figure*}

\section{Upper Limits and kilonova constraints} \label{sec:model}

\subsection{Upper Limits and Completeness} \label{subsec:upperLimit}

Figure~\ref{fig:limmag} shows the 5$\sigma$ limiting magnitudes we measured on the difference images for SOAR, Wendelstein, DECam and T80S observations. We employ these measurements to derive constraints on kilonova parameters using both BNS and BHNS kilonova models.
Based on the three-dimensional localization probability using W1-band luminosity from \citet{NED6_GCN}, we performed an approximate estimate to assess the likelihood that our galaxy-targeted observations covered the true host of S250206dm. The SOAR observations cover approximately 1.3$\%$ of the probability, and the Wendelstein 3KK observations about 1.5$\%$.
The estimation is performed against a luminosity function, capturing a true completeness including the 3D localization of the individual galaxies (Gassert et al., in prep.). It is based on the idea of using luminosities (here in $r$-band/W1-band) as a proxy for stellar mass. By using a luminosity function as expectation and comparing the true luminosities of the individual galaxies in the target catalog we estimate the completeness in the specific 3D volume of the GW event. This approach incorporates not only the luminosity function as is, but weighs the contributions based on their 3D localization such that galaxies with high luminosity in the outskirts of the localization volumes count less towards the completeness compared to similar galaxies (in terms of luminosity) in a high-probability region.
Complementing these targeted observations, the DECam observations covered a broader region, comprising 9.3 $\%$ of the cumulative localization probability in the LVK skymap.
T80S observations cover 4.9 $\%$ of the cumulative localization probability from the LVK skymap, with 4.5 $\%$ overlapping the DECam footprint.

\subsection{Kilonova models} \label{subsec:models}

We use state-of-the-art kilonova models computed with the 3D Monte Carlo radiative transfer code \textsc{possis} \citep{Bulla2019,Bulla2023}, as also done in Ahumada et al. (in prep.). In summary, ejecta are modeled under the assumption of axial symmetry and are composed of two components \citep{Nakar2020}: the first is ejected on dynamical timescales during the merger (``dynamical ejecta''), while the second is launched afterward from an accretion disk formed around the merger remnant (``disk wind ejecta''). Angular profiles for the ejecta density and composition are implemented following suggestions from numerical relativity simulations, which lead to a viewing angle dependence of the signal. The BNS kilonova dynamical ejecta follow a density $\rho\propto \sin^2\theta$ and an electron fraction $Y_{\rm e}(\theta)\propto \cos^2 \theta$ profile, where $\theta$ is the polar angle with respect to the binary angular momentum, while the wind ejecta have uniform $Y_{\rm e}$ and spherically symmetric density \citep{Perego2017,Radice2018,Setzer2023}. For what concerns the BHNS models, the dynamical ejecta are more strongly focused around the merger plane -- $\rho\propto 1/\{1+\exp[-20\,(\theta-1.2)]\}$ -- and their electron fraction is fixed to $Y_{\rm e}=0.1$, whereas the wind ejecta are spherical and have a flat distribution of $Y_{\rm e}$ values between 0.2 and 0.3 \citep[][and reference therein]{Kawaguchi_2020}. The viewing angle dependence is explored by computing kilonova observables for 11 different viewing angles equally spaced in cosine from the merger plane ($\cos\theta_{\rm obs}=0$, edge-on case) to the pole ($\cos\theta_{\rm obs}=1$, face-on). The heating rates are updated compared to previous work and consistent with those from \citet{Rosswog2024}.
 
Inspired by \cite{Anand2023}, a grid of 3072 BNS kilonova models is generated by varying the mass, velocity, and electron fraction of both ejecta components within ranges suggested by numerical relativity simulations: dynamical ejecta mass $M_{\rm dyn}=(0.001,0.005,0.01,0.02)\,M_{\odot}$, mass-weighted dynamical ejecta velocity $\bar{v}_{\rm dyn}=(0.12,0.15,0.2,0.25)\,$c, mass-weighted dynamical ejecta electron fraction  $\bar{Y}_{\rm e,dyn}=(0.15,0.20,0.25,0.30)$, wind ejecta mass $M_{\rm wind}=(0.01,0.05,0.09,0.13)\,M_{\odot}$, mass-weighted wind ejecta velocity $\bar{v}_{\rm wind}=(0.03,0.05,0.10,0.15)\,$c, and fixed wind ejecta electron fraction $Y_{\rm e,wind}=(0.20,0.30,0.40)$. 
 
Following \cite{Mathias2024}, a separate grid of 108 BHNS kilonova models is constructed by varying the masses of the two compact objects, the black hole spin, and the equation of state, with the following choices: $M_{\rm NS}=(1.2,1.4,1.6,1.8)\,M_{\odot}$, $M_{\rm BH}=(4.0,6.0,8.0)\,M_{\odot}$, $\chi_{\rm BH}=(0.0,0.3,0.6)$, and EOS (from stiff to soft)=(DD2, AP3, SFHo+H$\Delta$) \citep{Hempel2010,Akmal1998,Drago2014}. Numerical relativity fitting formulae are used to determine the properties of the ejecta, with only 37 out of 108 BHNS models leading to material being ejected and therefore resulting in a kilonova signal.  When taking into account the viewing angle as an additional free parameter, the total number of simulated kilonovae in the two grids is 33\,792 and 407, respectively.

A uniform grid of $\cos\theta_{\rm obs}$ is adopted to account for different viewing angles for both BHNS and BNS models. However, this set-up does not reflect the expected probability distribution of viewing angles for such binary systems. 
Given the absence of specific constraints on the inclination with respect to the line of sight for this event, here we incorporate a prior distribution of viewing angles of compact binary coalescences introduced by \citet{Schutz_2011}, which peaks at $\sim 30^\circ$. We assign to each model a weight corresponding to the probability of its viewing angle under the Schutz distribution. Throughout this paper, all reported fractions of model constraints are weighted accordingly to account for this effect.

\subsection{Kilonova constraints} \label{subsec:modeConstrain}

In Fig. \ref{fig:lightcurves} we show lightcurves from our model grids (across different viewing angles) at a distance of 323 Mpc, the mean of the luminosity distance posterior distribution over the southern sky lobe of high probability covered by DECam and SOAR. The 1$\sigma$ luminosity distance constraint around this median is 82 Mpc in this region. 
We compare these lightcurves with our upper limits, and show in color the lightcurves that are excluded by our non-detections.
Milky Way extinction at the location ($E(B-V) = 0.1 \pm 0.02$ mag using \citealt{CCM89}) is accounted for in the observed upper limits before evaluating model viability.
%  We account for the Milky Way extinction at this location ($E(B-V) = 0.1 \pm 0.02$).
In the upper row, we show the BHNS kilonova models in $griJ$ bands. It is clear that the BHNS models can only be constrained by the SOAR $g$ band observations, and marginally by the SOAR $i$ band. Overall, $15\%$ of the models are disfavored by the SOAR data. Had the DECam observations been possible at the same time as the SOAR ones instead of 6 days later, they would have ruled out  $51\%$ of BHNS models over a significantly larger high probability area, showing the importance of early observations.

In the middle and lower rows of Fig. \ref{fig:lightcurves} we show the BNS kilonova lightcurves along with those ruled out by SOAR and DECam observations respectively. A larger fraction of models can be ruled out in this case by both follow up campaigns. Again, the early SOAR $g$ band observations exclude the largest fraction of kilonova models ($55\%$), though over a much smaller area than the DECam observations, which in this case constrain $1.4\%$ of the kilonova models thanks to the deep $i$ band pointings. 
The models excluded by the DECam follow up comprise 43\% of all the kilonova models at least as bright as AT 2017gfo, the kilonova associated to GW170817, within the first week after merger\footnote{The time window is chosen to match the timing of the SOAR observations and the first (most constraining) DECam epoch were conducted within the first week post-merger. Beyond this period, an AT 2017gfo-like kilonova at 323 Mpc typically fades too much in optical bands to be detectable by ground-based telescopes.}, whereas SOAR observations rule out the full set of such kilonova models.
At 6 days post-merger, the kilonova is expected to be redder than earlier epochs, e.g., AT 2017gfo had an $r-i$ color of $\sim 0.10$ mag at 0.5 days and $\sim 0.76$ mag at 6 days \citep{2017ApJ...851L..21V}, which coupled to our deeper limits in $i$ compared to $r$, rendered our $i$ band depths more constraining. DECam + T80S jointly rule out 4\% of the kilonova models, which comprise 48\% of all the kilonova models at least as bright as AT 2017gfo.

The Wendelstein observations, which were galaxy-targeted over the northern high-probability region of the sky map, correspond to a luminosity distance of $383 \pm 96$ Mpc. 
Assuming the event occurred at the optimistic closest $1\sigma$ end of the distance distribution, i.e. 287\,Mpc, the Wendelstein data would marginally constrain kilonova models - ruling out the brightest $\sim 0.1\%$ models at $\sim1$ day post-merger.

\begin{figure*}
\centering
\gridline{
  \fig{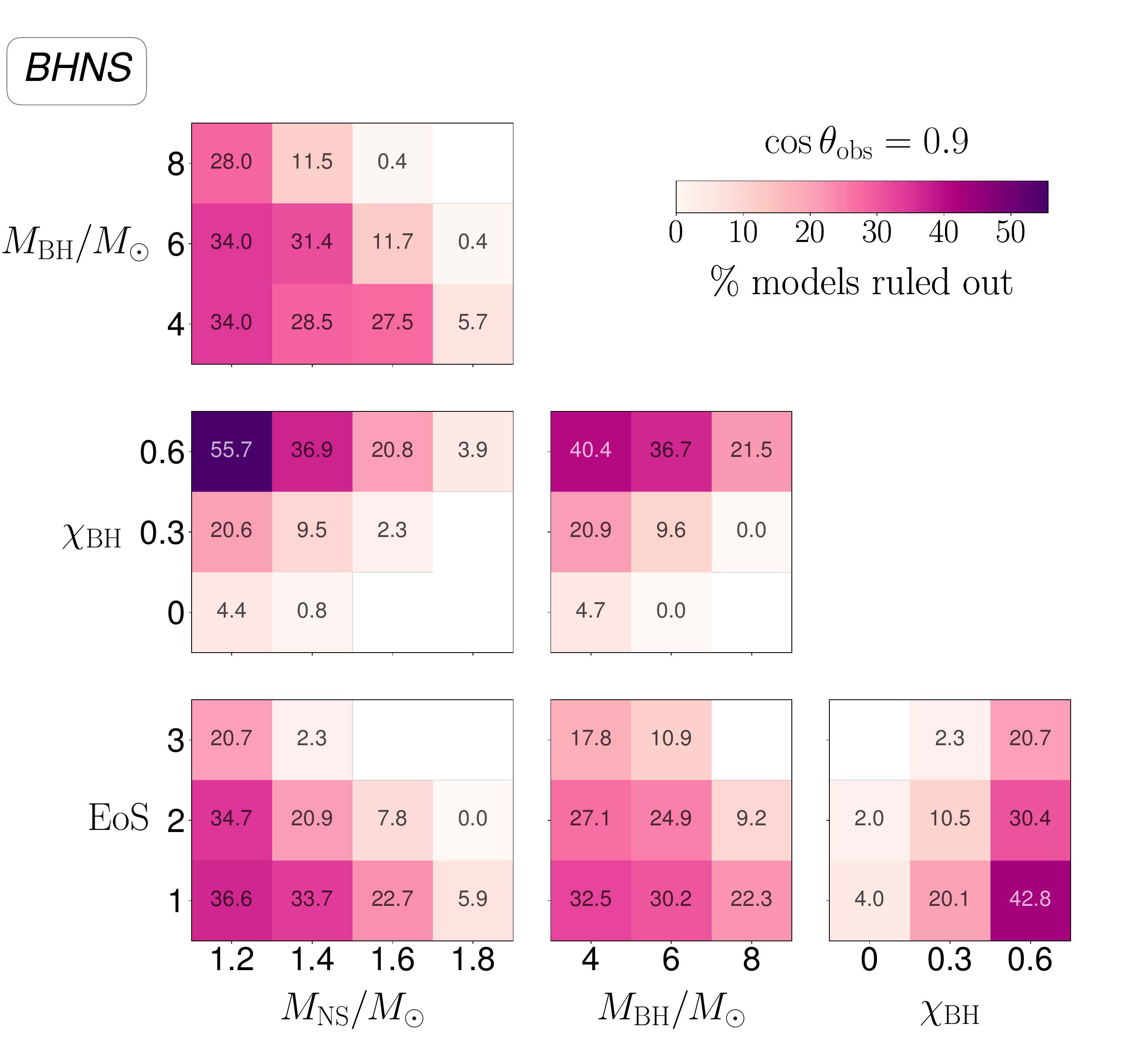}{0.49\textwidth}{(a) BHNS model constrained by SOAR observations}
  \fig{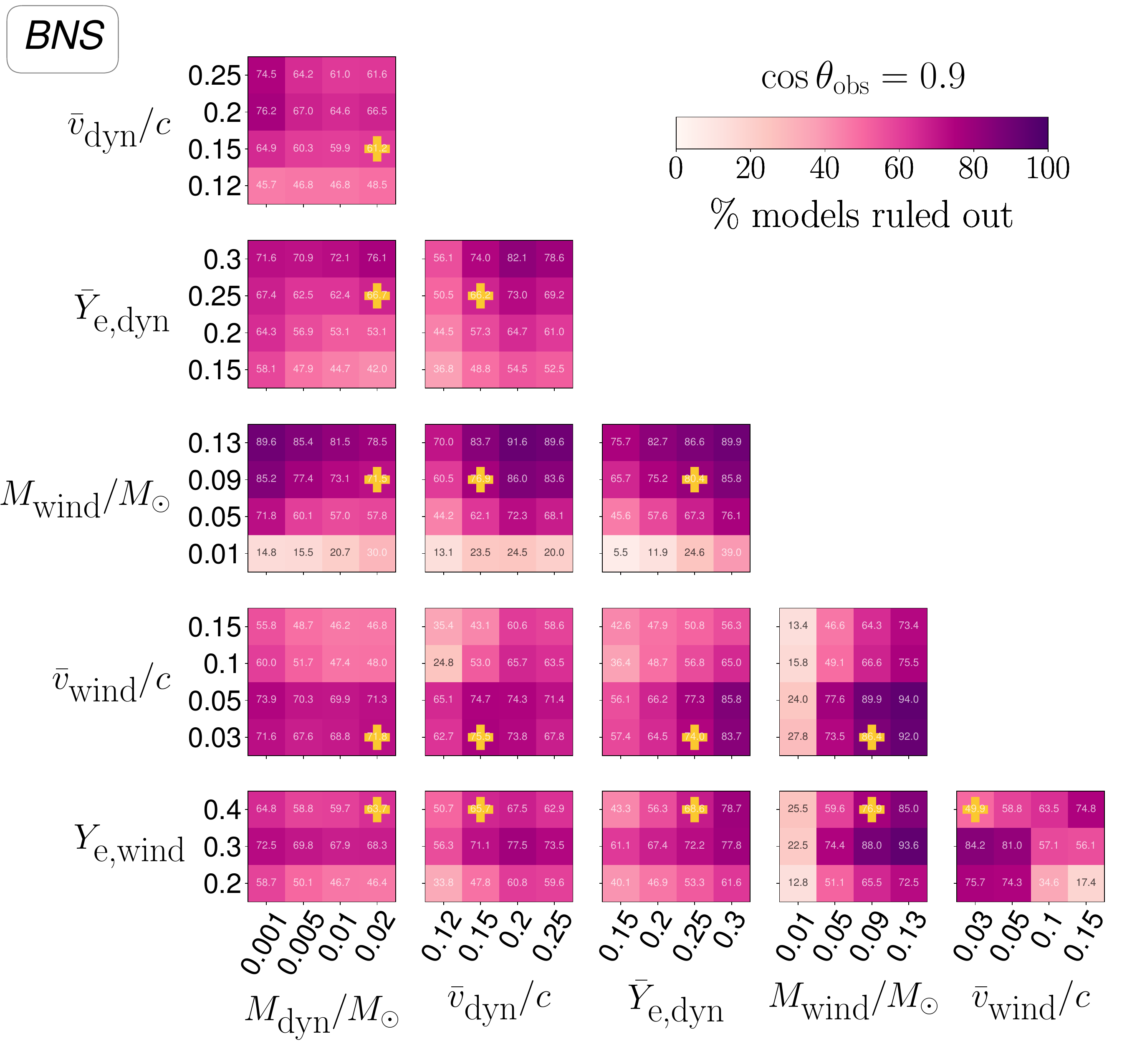}{0.49\textwidth}{(b) BNS model constrained by SOAR observations}
}
\gridline{
  \fig{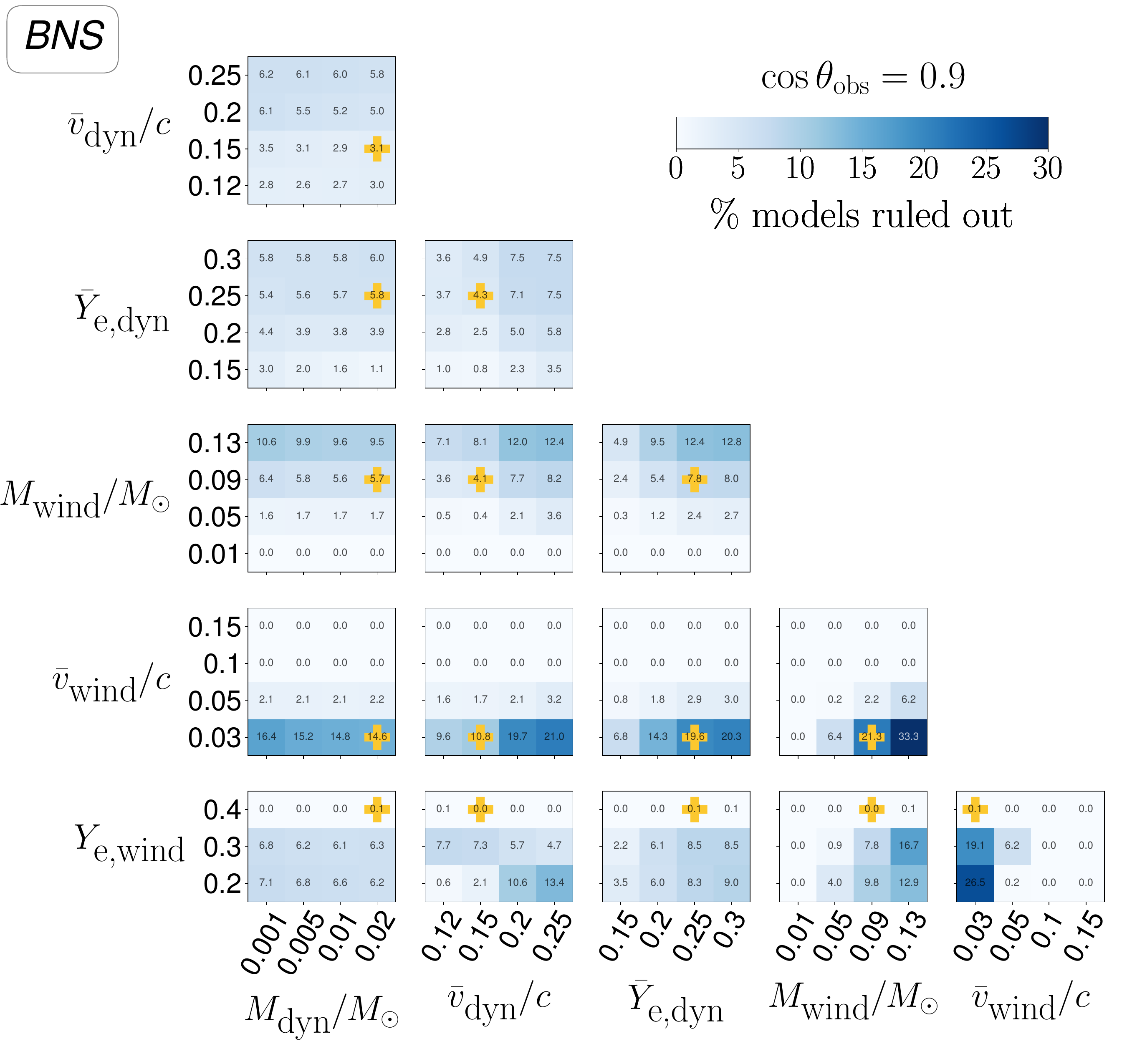}{0.49\textwidth}{(c) BNS model constrained by DECam observations}
  \fig{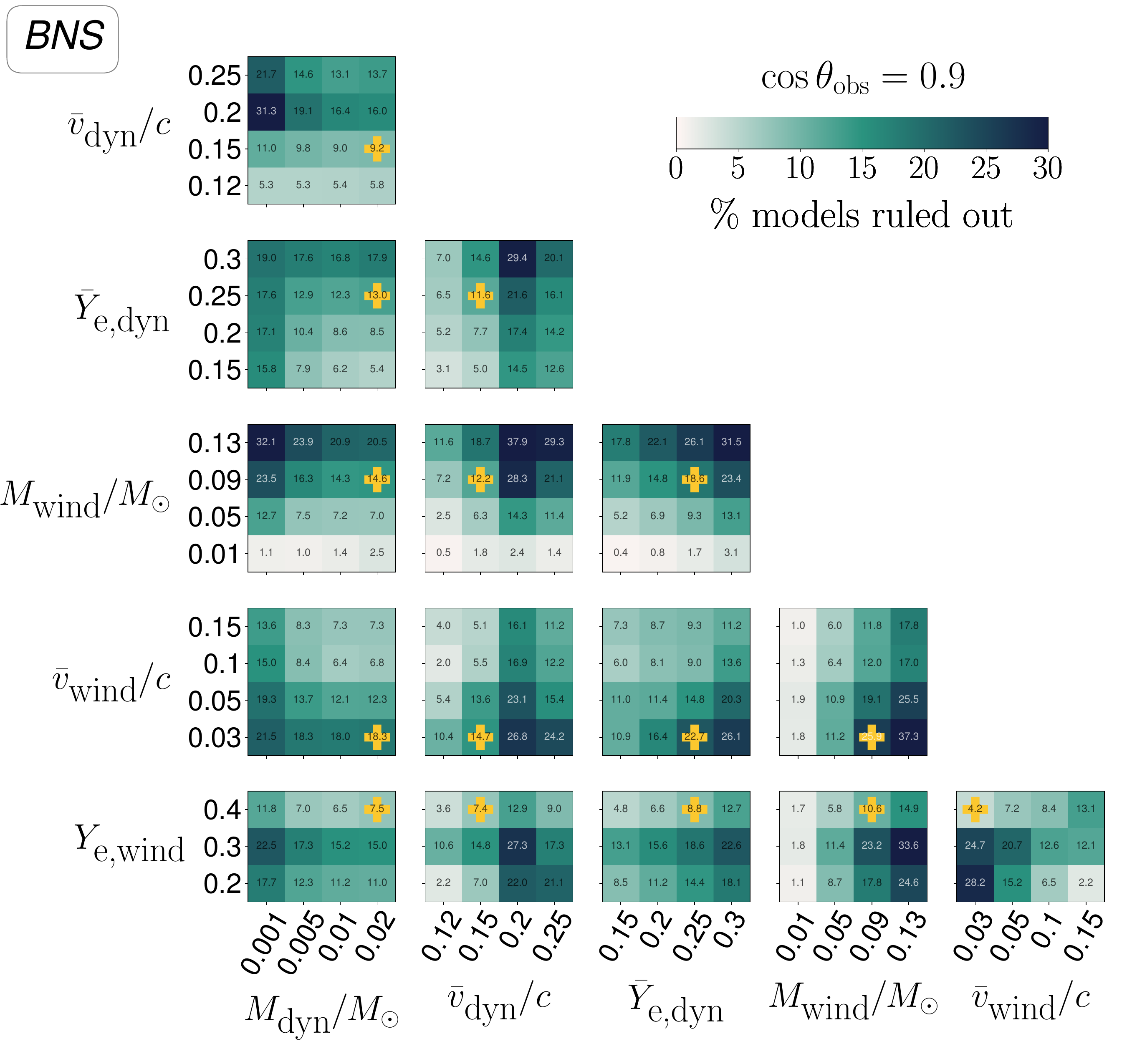}{0.49\textwidth}{(d) BNS model constrained by DECam and T80S observations}
}
\caption{Corner plot showing the fraction of BHNS and BNS models ruled out by our observations with a viewing angle $\theta$ with ${\rm cos}(\theta) = 0.9$, represented by a color bar. The number labeled on each cell represents the overall exclusion fraction (\%), conditional on given model parameters, computed by averaging over sampled distances. For BNS models, the yellow cross marker indicates the best-fit parameters of AT2017gfo from Farinelli et al. (submitted).}
\label{fig:model_constrain}
\end{figure*}

In Fig.~\ref{fig:model_constrain} we show how the excluded models translate into constraints on physical parameters. As our fiducial scenario, we consider the case where the viewing angle is off-axis with ${\rm cos}(\theta) = 0.9$ (the grid point nearest to the peak of the Schutz distribution of viewing angles), and present the percentage of models ruled out in each parameter combination. The value of ${\rm cos}(\theta) = 0.9$ corresponds to a viewing angle of $\sim 26$ deg, which is close to both the GW170817 viewing angle (e.g. \citealt{Mooley_2018,2021ARA&A..59..155M,Palmese_2024}) and the expected peak of the distribution of angles from an observed population of GW events \citep{Schutz_2011}. To account for the uncertainty in the event's distance, we sample 256 values from the posterior distribution of the luminosity distance over the southern sky lobe of high probability (with a mean distance of 323 Mpc). The exclusion fraction is computed at each sampled distance and then averaged to yield the overall constraint illustrated in Fig.~\ref{fig:model_constrain}.
The upper panel shows the constraints from SOAR on the BHNS models. EoS numbers 1, 2, and 3 correspond to  DD2 \citep{Hempel2010}, AP3 \citep{Akmal1998}, and SFHo+HDelta \citep{Drago2014} respectively, going from stiffer to softer. As expected, the largest fraction of models ruled out lies in the lowest NS mass bin and stiffer EoS, since in this case the neutron star is expected to be more easily disrupted by the black hole, producing a brighter counterpart. With a DD2 EoS and a $\sim 1.2 ~M_\odot$ neutron star, about $37\%$ of our models are excluded. 
%From 30 to more than $50\%$...
From $21\%$ to $56\%$ of our models are excluded for our highest spinning black holes, at a 0.6 dimensionless spin magnitude, and a $1.2\mbox{--}1.6~M_\odot$ neutron star. This is due to the Innermost Circular Orbit (ISCO) radius reduction for highly spinning black holes compared to the non-spinning case, leading to more favorable disruption of the neutron star around it. 
About $0-39\%$ (where the range is due to the EOS considered; if one accounts for all EOSs then this number is 28\%) of models with black holes in the mass gap are also excluded, which is a larger fraction than those at higher masses, as they will also have a larger ISCO.

For what concerns the BNS kilonova models, shown in the lower panels of Fig. \ref{fig:model_constrain}, the SOAR observations rule out a large fraction, about $90\%$ of massive ($M_{\rm wind} \gtrsim 0.09 M_\odot$) slow winds ($\bar{v}_{\rm wind}<0.1c$). This is because, as expected, the larger wind ejecta mass gives rise to the brighter kilonovae that we can exclude, while the slower wind ejecta give rise to the longer-lasting emission. Large electron fractions are also significantly excluded by our observations ($70\%$ of models with $\bar{Y}_{\rm e,dyn}\gtrsim 0.2$ and $Y_{\rm e,wind}\gtrsim0.3$). These models are in fact expected to give rise to blue, short-lived kilonovae that are well constrained by the early deep SOAR $g$ band limits. 

The DECam observations also rule out a significant fraction of massive ($M_{\rm wind} \gtrsim 0.09 M_\odot$), slow ($\bar{v}_{\rm wind}\sim0.03c$) wind ejecta, albeit for a smaller number of models ($\sim 30\%$ for DECam and DECam+T80S).  We also find a slight trend where a larger fraction of models are ruled out at lower dynamical ejecta masses ($M_{\rm dyn}\lesssim 0.001~M_\odot$) compared to higher masses, as smaller dynamical ejecta lead to a lower ``shielding'' effect the inner wind ejecta and hence a brighter kilonova \citep{Kawaguchi_2020}. This is also observed at the larger viewing angles explored in this work.
As opposed to the SOAR case, the DECam limits do not have significant constraining power on the highest wind electron fraction $Y_{\rm e,wind}\sim0.4$ (ruling out less than $1 \%$ of models since they lack the early blue observations), but they are relatively more constraining at $Y_{\rm e,wind}\lesssim0.3$ (rejecting $\sim 7\%$) thanks to the late $i$ band observations. Combining DECam and T80S increases the exclusion fractions to 8\% for \( Y_{\rm e,wind} \sim 0.4 \) and 15\% for \( Y_{\rm e,wind} \lesssim 0.3 \), respectively.
We do not show kilonova constraints for the Wendelstein observations as they can only constrain a few models at the low-end tail of the luminosity distance posterior due to the shallow depths reached in poor observing conditions.

As expected, the BHNS models we consider are fainter and therefore a smaller fraction is ruled out by our observations compared to the BNS ones. The BHNS kilonovae used here go down to ejecta masses of $10^{-4}~M_\odot$ \citep{Mathias2024}, which is about an order of magnitude below the lowest value we assume for BNSs. Moreover, the BHNS kilonova is expected to lack a polar dynamical ejecta component, as reflected by our models, so an on axis observer is more likely to be able to observe a brighter BNS than NSBH kilonova, all kilonova parameters but the squeezed polar component being equal. 

Results from our BNS model grid can be compared to the work of \citet{anand2021}, which derived kilonova constraints for BHNS mergers in O3. Overall, their constraints on S200115j (GW200115\_042309) are less stringent than ours as only some massive ejecta ($>0.05~M_\odot$) models are ruled out at the lower end of the luminosity distance estimate, while S200105ae (GW200105\_162426, later deemed a sub-threshold event) follow-up did not allow for any constraint of kilonova models. A major difference between the two analyses consists in the grid values, for which the ejecta masses (in the range $0.01-0.09~M_\odot$) are higher than in our models ($0.001-0.02~M_\odot$), therefore giving rise to brighter EM counterparts. Bright kilonovae for BHNS mergers (at least compared to BNS mergers) are motivated by previous findings (e.g. \citealt{Rosswog_2017,Barbieri_2019,Kawaguchi_2020}), however, this is highly dependent on the components' mass, spin, and neutron star EOS. More stringent constraints were possible on GW190814 using DECam observations \citep{2020ApJ...890..131A,Morgan_2020,anand2021}, thanks to the availability of early observations and the small localization are of the event. However, GW190814 is likely a binary black hole merger, thus it may have not produced a kilonova \citep{GW190814,Tews2021}. 

The angular dependence of the models we use is such that the most stringent constraints are obtained for on-axis systems. Since the lanthanide-rich ejecta are launched around the plane of the binary within some angle, emission from kilonovae observed edge-on will face a larger optical depth than those observed on-axis, and will therefore appear significantly fainter \citep[e.g.,][]{Bulla2019,Darbha2020,Korobkin2021,Mathias2024}. We show kilonova constraints at different viewing angles in Appendix~\ref{Appendix:Constrain_viewAngle}. We note that results in the cos$(\theta_{\rm obs})=0.9$ scenario are qualitatively similar to but less stringent (in terms of model constraints) than the on-axis case. 

\section{Discussion and Conclusions} \label{sec:conclusion}

We have presented our observational campaign aimed at identifying a kilonova associated with the high-significance gravitational-wave event candidate S250206dm, which is a probable BHNS or BNS merger. In the BHNS case, this event is interesting in light of the possibility that the black hole had a low mass, $3\mbox{--}5~M_\odot$, which would place it within the mass gap as well as increasing the likelihood of an EM counterpart as compared to other BHNS systems. This event likely represents the most exciting event of LIGO/Virgo/KAGRA O4 so far for multimessenger searches, given its low False Alarm Rate and significant probability of having ejected matter outside of the remnant ISCO.

Under the assumption of an off-axis viewing angle with ${\rm cos}(\theta) = 0.9$,
assuming that this event is a mass gap BHNS, we rule out up to $\sim40\%$ of simulated models conditional on the neutron star mass and EoS around the top 33 host galaxies in the Southern lobe of this event's localization. Assuming BNS kilonovae described by \citet{Bulla2023} instead, we are able to rule out about $90\%$ of models for those galaxies with massive ($\gtrsim 0.09~M_\odot$) and slow ($\bar{v}_{\rm wind}<0.1c$) wind ejecta. $74\%$ of models with wind ejecta mass $\gtrsim0.05~M_\odot$ are also ruled out. 

Our DECam (DECam + T80S) constraints on the other hand cover $\sim 9.3\%$ (4.5\%) of the sky probability and constrain  $\sim$ 43\% (48\%) of BNS models brighter or similar to the GW170817 kilonova. 
Combined with ZTF limits, kilonova models brighter than $-15$ mag that rise (with evolution rate $\alpha <$ 0 mag/day), as well as those brighter than -16.5 mag slowly fading ($\alpha  >$ 0.3 mag/day), can be ruled out over the majority ($77\%$) of the sky probability of this event (Ahumada et al., in prep.). If DECam observations had been allowed on the day of trigger, and a similar depth was reached, 51\% of our BHNS and 82\% of our BNS models would have been constrained. This work shows the potential for detection of kilonovae associated with both BNS and NSBH mergers at distances $>300~$Mpc, provided that deep and prompt observations can be carried out.

%% Please use the acknowledgment and contribution environments. This will 
%% be anonomyized when the "anonymous" style option is used. 

\begin{center}
\section*{Acknowledgments}
\end{center}
\vspace{-1em}
% \begin{acknowledgments}
LH, TC, and AP acknowledge that this material is based upon work supported by NSF Grant No. 2308193. BO gratefully acknowledges support from the McWilliams Postdoctoral Fellowship at Carnegie Mellon University. MB acknowledges the Department of Physics and Earth Science of the University of Ferrara for the financial support through the FIRD 2024 grant. IA acknowledges support from the National Science Foundation Award AST 2505775 and NASA grant 24-ADAP24-0159. This work was supported by the Deutsche Forschungsgemeinschaft (DFG, German Research Foundation) under Germany's Excellence Strategy – EXC-2094 – 390783311. M.W.C. acknowledges support from the National Science Foundation with grant numbers PHY-2117997, PHY-2308862 and PHY-2409481.

This work used resources on the Vera Cluster at the Pittsburgh Supercomputing Center (PSC). We thank the PSC staff for help with setting up our software on the Vera Cluster.

This project used data obtained with the Dark Energy Camera (DECam), which was constructed by the Dark Energy Survey (DES) collaboration.
Funding for the DES Projects has been provided by the US Department of Energy, the US National Science Foundation, the Ministry of Science and Education of Spain, the Science and Technology Facilities Council of the United Kingdom, the Higher Education Funding Council for England, the National Center for Supercomputing Applications at the University of Illinois at Urbana-Champaign, the Kavli Institute for Cosmological Physics at the University of Chicago, Center for Cosmology and Astro-Particle Physics at the Ohio State University, the Mitchell Institute for Fundamental Physics and Astronomy at Texas A\&M University, Financiadora de Estudos e Projetos, Fundação Carlos Chagas Filho de Amparo à Pesquisa do Estado do Rio de Janeiro, Conselho Nacional de Desenvolvimento Científico e Tecnológico and the Ministério da Ciência, Tecnologia e Inovação, the Deutsche Forschungsgemeinschaft and the Collaborating Institutions in the Dark Energy Survey.

The Collaborating Institutions are Argonne National Laboratory, the University of California at Santa Cruz, the University of Cambridge, Centro de Investigaciones En\`ergeticas, Medioambientales y Tecnol\`ogicas–Madrid, the University of Chicago, University College London, the DES-Brazil Consortium, the University of Edinburgh, the Eidgenössische Technische Hochschule (ETH) Zürich, Fermi National Accelerator Laboratory, the University of Illinois at Urbana-Champaign, the Institut de Ci\'encies de l’Espai (IEEC/CSIC), the Institut de F\'isica d’Altes Energies, Lawrence Berkeley National Laboratory, the Ludwig-Maximilians Universit\:at M\:unchen and the associated Excellence Cluster Universe, the University of Michigan, NSF’s NOIRLab, the University of Nottingham, the Ohio State University, the OzDES Membership Consortium, the University of Pennsylvania, the University of Portsmouth, SLAC National Accelerator Laboratory, Stanford University, the University of Sussex, and Texas A\&M University.

Based on observations at Cerro Tololo Inter-American Observatory, NSF’s NOIRLab (NOIRLab Prop. ID 2023B-851374, PI: Andreoni \& Palmese), which is managed by the Association of Universities for Research in Astronomy (AURA) under a cooperative agreement with the National Science Foundation.
We thank Kathy Vivas, Alfredo Zenteno, and CTIO staff for their support with DECam observations. Some of the observations reported in this paper were obtained with the Southern African Large Telescope (SALT).

This research has made use of the NASA/IPAC Extragalactic Database (NED), which is funded by the National Aeronautics and Space Administration and operated by the California Institute of Technology.

\facilities{CTIO:4m, SOAR, FTW:2.1m, SALT}


\software{
    Astropy \citep{astropy13,astropy18,astropy22}, 
    SciPy \citep{2020SciPy-NMeth},
    Numpy \citep{2020NumPy-Array},
    MatPlotLib \citep{Matplotlib},
    CuPy \citep{CuPy},
    Scikit-image \citep{skimage},
    SourceExtractor \citep{SExtractor},
    SWarp \citep{SWarp},
    SFFT \citep{sfft_zenodo},
    dustmap \citep{Green2018},
    ligo.skymap \citep{Singer_2016,Singer_2016_supp},
    healpy \citep{Zonca2019},
    Treasure Map \citep{Wyatt_2020}
}


\clearpage

\appendix

\section{Candidates from DECam Search} \label{Appendix:DECam_Candidates}

Table~\ref{table:DECam_candidates} lists the candidates that passed visual inspection from the DECam search. Most of the candidates were excluded because of their slow photometric evolution during our observations (i.e. $<$ 0.3 mag/day). Of the two remaining sources, AT 2025me had previous variability, which was reported prior to the event S250206dm; AT 2025bno is a hostless transient rejected by its blue color.

\begin{table*}
    \centering
    \caption{Candidates from DECam search.}
    \label{table:DECam_candidates}
    \begin{tabular}{lllllll}%|l|}
    \hline\hline 
    Candidate & Coordinates (R.A., Decl.) & Discov. Mag. & Discov. Color Mag. & Phot. Evol. & Redshift & Comment \\ \hline
    \multicolumn{6}{c}{\textit{Fast Decline}} \\
    AT 2025bno  & 16:10:47.65 $-$68:28:13.6 & $i$ = 21.10 & $r - i$ = -0.18 (-0.23) & $\geqslant$ 0.52 ($i$) &... & Hostless \\% \hline
    AT 2025me  & 16:49:43.68 $-$70:23:34.2 & $r$ = 18.90 & $r - i$ = -0.11 (-0.16) & 0.94 ($r$) &... & Pre-event detection \\% \hline
    \multicolumn{6}{c}{\textit{Flat or Slowly Decline}} \\
    AT 2025bmy  & 15:42:37.02 $-$69:36:49.6 & $i$ = 21.28 & $r - i$ = 0.97 (0.92)   & 0.07 ($i$)  &... &... \\% \hline
    AT 2025bnh  & 16:32:38.20 $-$68:31:02.9 & $r$ = 22.54 & $r - i$ = 0.96 (0.90)   & 0.11 ($r$)  & 0.140$\,_\texttt{SALT}$  & Nuclear \\% \hline
    AT 2025bmw  & 16:18:12.69 $-$71:56:17.4 & $r$ = 21.24 & $r - i$ = 0.88 (0.83)   & 0.09 ($r$)  &... & Nuclear \\% \hline
    AT 2025bnb  & 15:50:13.08 $-$70:19:04.8 & $i$ = 20.87 & $r - i$ = 0.80 (0.72)   & 0.04 ($i$)  &0.117$\,_\texttt{NED}$ &... \\% \hline
    AT 2025bnt  & 16:33:49.52 $-$70:02:56.8 & $i$ = 21.36 & $r - i$ = 0.78 (0.73)   & 0.10 ($r$)  &... &... \\% \hline
    AT 2025bmx  & 15:57:20.48 $-$68:40:02.4 & $i$ = 21.40 & $r - i$ = 0.72 (0.66)   & 0.07 ($i$)  &... & Nuclear \\% \hline
    AT 2025btj  & 16:06:31.80 $-$66:57:35.3 & $i$ = 22.06 & $r - i$ = 0.65 (0.59)   & 0.02 ($i$)  &... & Hostless \\% \hline
    AT 2025bts  & 16:45:09.06 $-$71:31:23.2 & $i$ = 22.61 & $r - i$ = 0.64 (0.59)   & 0.04 ($i$)  &... &... \\% \hline
    AT 2025bnu  & 16:51:27.56 $-$69:06:28.5 & $r$ = 22.04 & $r - i$ = 0.59 (0.54)   & 0.21 ($r$)  &... & Nuclear \\% \hline
    AT 2025bnp  & 15:52:22.31 $-$70:04:42.6 & $i$ = 21.68 & $r - i$ = 0.53 (0.46)   & -0.02 ($i$) & 0.100$\,_\texttt{SALT}$  &... \\% \hline
    AT 2025fso  & 15:30:50.47 $-$70:27:21.7 & $i$ = 21.99 & $r - i$ = 0.51 (0.44)   & 0.24 ($r$)  &... &... \\% \hline
    AT 2025bnc  & 17:05:02.58 $-$69:52:26.0 & $r$ = 22.62 & $r - i$ = 0.49 (0.43)   & 0.06 ($i$)  &... & Hostless \\% \hline
    AT 2025fsy  & 16:26:53.13 $-$70:00:47.0 & $i$ = 21.89 & $r - i$ = 0.47 (0.41)   & 0.05 ($r$)  &... & Nuclear \\% \hline
    AT 2025bms  & 16:28:24.40 $-$70:48:57.9 & $r$ = 22.13 & $r - i$ = 0.42 (0.37)   & 0.23 ($r$)  &... & Nuclear \\% \hline
    AT 2025fsq  & 15:57:15.75 $-$67:45:22.5 & $i$ = 22.48 & $r - i$ = 0.41 (0.35)   & 0.09 ($i$)  &... & Nuclear \\% \hline
    AT 2025bns  & 16:52:25.99 $-$69:04:18.6 & $r$ = 22.84 & $r - i$ = 0.39 (0.34)   & 0.09 ($r$)  &... &... \\% \hline
    AT 2025bni  & 16:33:59.30 $-$69:18:09.3 & $r$ = 22.58 & $r - i$ = 0.34 (0.27)   & 0.08 ($r$)  &... & Nuclear \\% \hline
    AT 2025fsx  & 15:43:23.69 $-$68:44:52.3 & $i$ = 22.48 & $r - i$ = 0.30 (0.24)   & -0.03 ($i$) &... &... \\% \hline
    AT 2025bth  & 16:52:03.71 $-$68:14:22.3 & $r$ = 20.40 & $r - i$ = 0.14 (0.10)   & 0.10 ($i$)  & 0.102$\,_\texttt{NED}$  &... \\% \hline
    AT 2025bnj  & 15:59:51.07 $-$66:55:46.9 & $i$ = 21.68 & $r - i$ = 0.06 (0.01)   & 0.01 ($i$)  & 0.140$\,_\texttt{SALT}$ &... \\% \hline
    AT 2025fsz  & 16:32:30.65 $-$68:42:58.8 & $r$ = 23.23 & $r - i$ = 0.00 (-0.05)  & 0.04 ($i$)  &... & Nuclear \\% \hline
    AT 2025fsw  & 15:47:15.50 $-$67:28:05.4 & $i$ = 22.39 & $r - i$ = -0.05 (-0.11) & 0.29 ($r$)  &... & Hostless \\% \hline
    AT 2025fst  & 16:32:39.71 $-$69:37:24.2 & $i$ = 23.09 & $r - i$ = -0.06 (-0.13) & 0.07 ($i$)  &... & Hostless \\% \hline
    AT 2025fsu  & 16:28:42.95 $-$71:20:20.6 & $r$ = 22.85 & $r - i$ = -0.12 (-0.17) & 0.01 ($r$)  &... &... \\% \hline
    AT 2025fta  & 16:11:52.15 $-$70:52:38.4 & $i$ = 22.82 & $r - i$ = -0.13 (-0.19) & 0.05 ($i$)  &... & Hostless \\% \hline
    AT 2025fsr  & 15:44:27.41 $-$68:10:13.6 & $i$ = 22.89 & $r - i$ = -0.13 (-0.19) & 0.01 ($r$)  &... &... \\% \hline
    AT 2025bnv  & 15:43:42.28 $-$67:40:57.0 & $i$ = 21.11 & $r - i$ = -0.15 (-0.22) & 0.08 ($i$)  &... &... \\% \hline
    AT 2025btc  & 16:06:40.98 $-$70:59:27.9 & $i$ = 22.14 & $r - i$ = -0.18 (-0.24) & 0.06 ($r$)  &... &... \\% \hline
    AT 2025fss  & 16:01:18.31 $-$70:59:41.1 & $i$ = 21.60 & $r - i$ = -0.21 (-0.27) & 0.01 ($i$)  &... &... \\% \hline
    AT 2025fsp  & 15:50:54.92 $-$70:26:25.3 & $i$ = 23.26 & $r - i$ = -0.26 (-0.34) & 0.28 ($r$)  &... &... \\% \hline
    AT 2024aexy & 16:07:47.10 $-$67:06:08.4 & $i$ = 21.23 & $r - i$ = -0.29 (-0.34) & 0.09 ($r$)  &... & Pre-event detection \\% \hline
    AT 2025bnm  & 16:22:52.18 $-$69:01:23.7 & $i$ = 22.04 & $r - i$ = -0.38 (-0.44) & 0.04 ($i$)  &... &... \\% \hline
    AT 2025bmz  & 16:40:26.39 $-$66:46:25.6 & $r$ = 19.83 & ...                     & 0.08 ($r$)  & 0.053$\,_\texttt{NED}$ & Nuclear \\% \hline
    AT 2025btr  & 16:58:02.83 $-$68:26:49.5 & $i$ = 22.29 & ...                     & 0.07 ($i$)  &... &... \\% \hline
    AT 2025fsv  & 16:48:20.86 $-$67:14:02.0 & $i$ = 22.60 & ...                     & 0.06 ($i$)  &... & Hostless \\% \hline
    \hline
    \end{tabular}
\end{table*}

\section{Spectroscopic Follow-up with SALT} \label{Appendix:SALT_Followup}

We obtained longslit optical spectroscopic observations of optical transients (and their host galaxies) potentially associated to S250206dm with the 10-m class Southern African Large Telescope \citep[SALT;][]{Buckley2006} at Sutherland Observatory in Sutherland, South Africa. Observations were obtained through a SALT Gravitational Wave Director's Discretionary Time proposal (ID: 2024-2-GWE-001; PI: T. Cabrera). 
Longslit spectroscopy of four candidates AT~20205bnt (2025-02-17); AT~2025bnj (2025-02-15 and 2025-02-16); AT~2025bnp (2025-02-14); and AT~2025bnh (2025-02-22) was acquired with the Robert Stobie Spectrograph \citep[RSS;][]{Burgh2003}. Due to the poor visibility of these fields to SALT at the time of the observations only a single target was able to be observed per night. Data was obtained between 2025-02-14 and 2025-02-22. The observation of AT~2025bnj was repeated as the initial data did not satisfy our observing condition requirements. 

Each spectrum was obtained using the PG0700 grating at a grating angle of $4.6^\circ$ (corresponding to a camera angle of $22.75^\circ$) with the PC03400 blocking filter using a slit width $1.5\arcsec$. The spectra covered the observer frame wavelength range 3592 \AA\ to 7479 \AA\ with resolution $R=735$ at the central wavelength of 5580\AA. We used an exposure time of $1\times900$ s per target. 

The data were reduced using the RSS Long-slit spectra processing and extraction app \texttt{rsslsspectra} (User Guide: \url{https://astronomers.salt.ac.za/wp-content/uploads/sites/71/2024/09/rsslsspectra.pdf}).
We extracted traces associated to the host galaxy of these transients.
We searched by eye for narrow emission features in the spectra for all targets, finding at least two regions associated with H$\alpha$, H$\beta$, O[II], or O[III] lines in each of the spectra of AT~2025bnj, AT~2025bnp, and AT~2025bnh, estimating best fit redshifts of $0.140 \pm 0.001$, $0.100 \pm 0.01$, and $0.140 \pm 0.001$, respectively. 
There were no discernible features in the spectrum for AT~20205bnt, and so we were unable to measure a redshift for this target. The results of SALT follow-up spectroscopy are summarized in Table~\ref{tab:salt}. 

\section{Kilonova constraints with different viewing angle} \label{Appendix:Constrain_viewAngle}

\begin{figure*}
\gridline{\fig{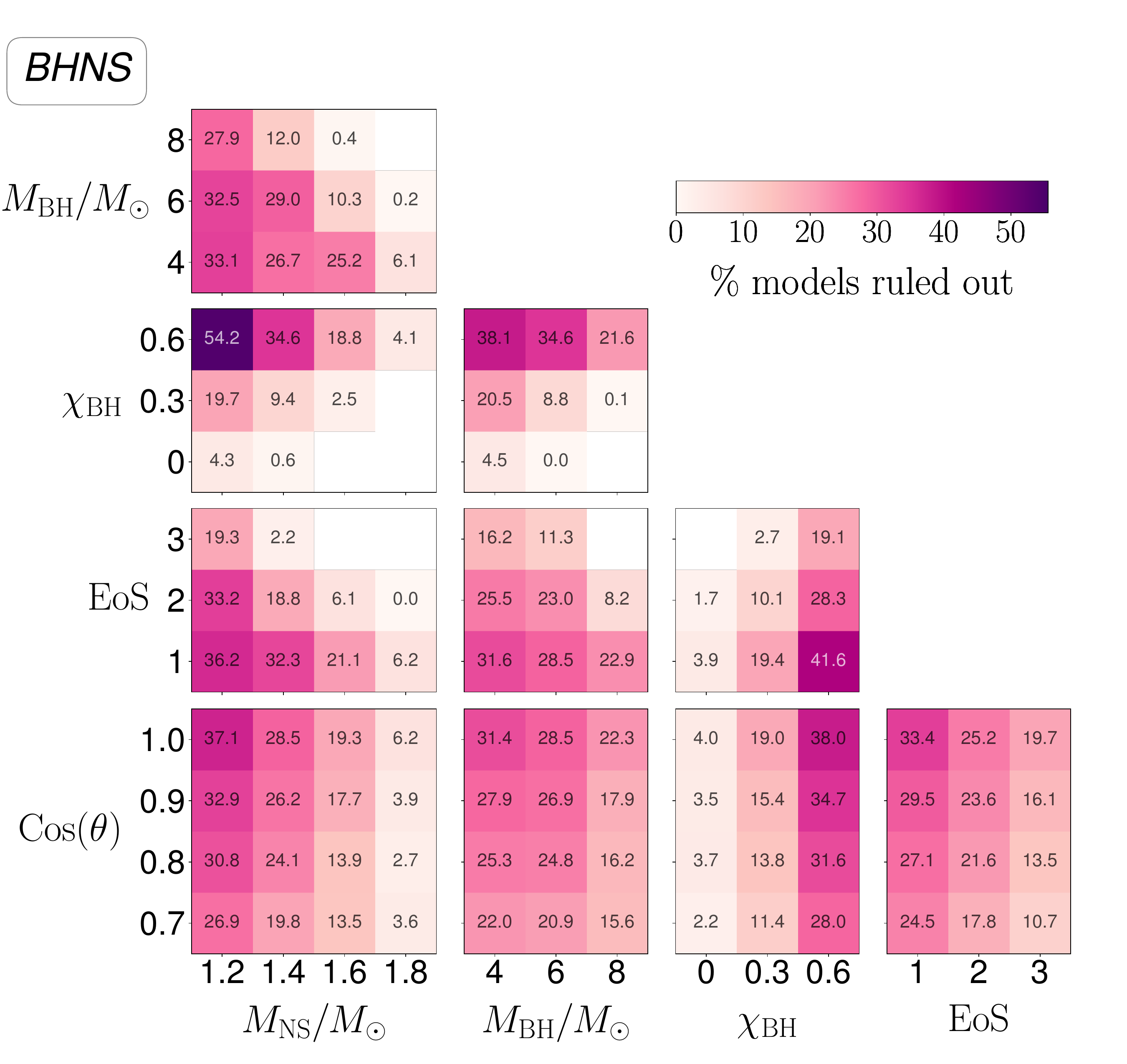}{0.48\textwidth}{(a) BHNS model constrained by SOAR observations}
          \fig{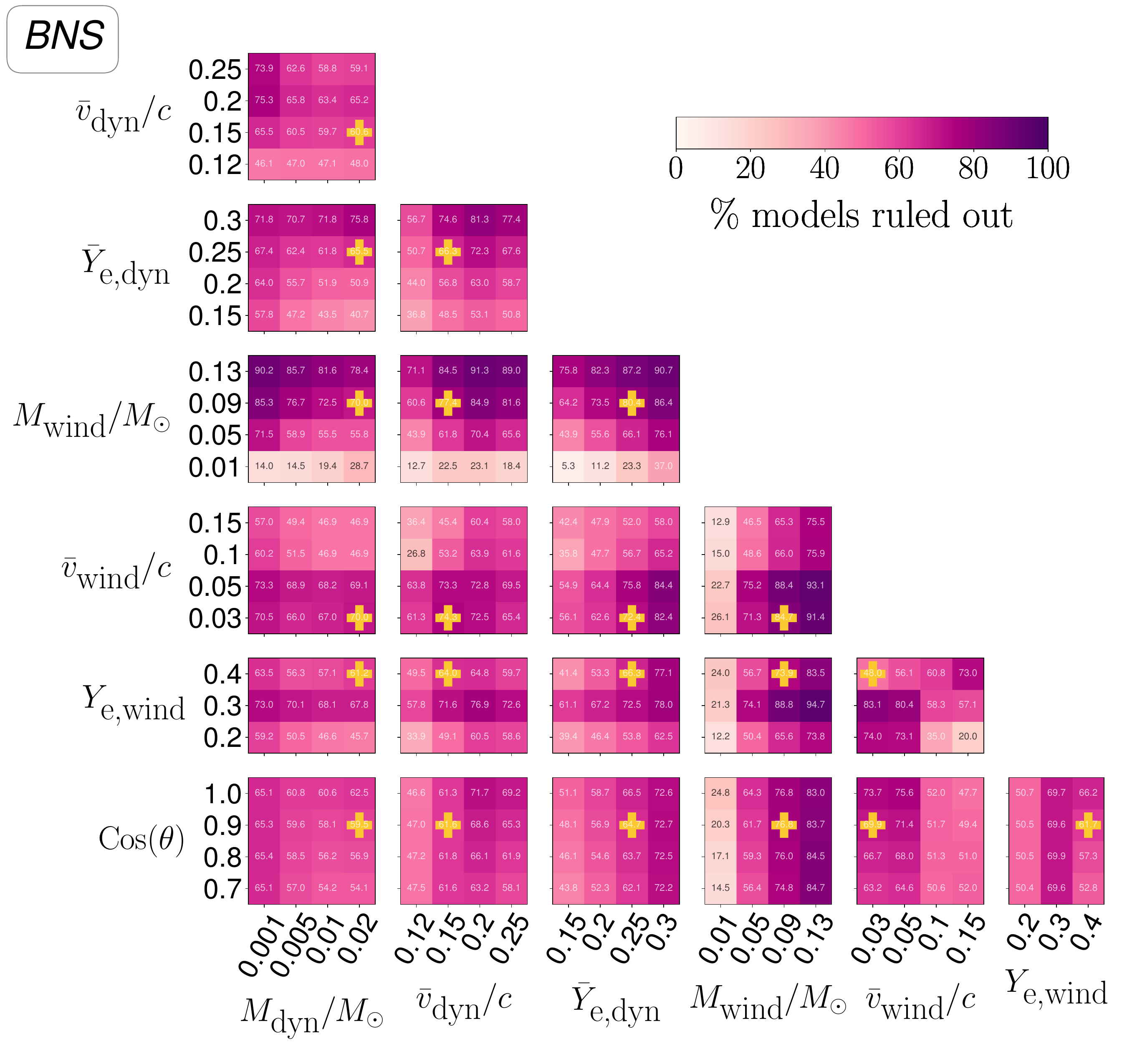}{0.48\textwidth}{(b) BNS model constrained by SOAR observations}}

\vspace{1em}

\gridline{\fig{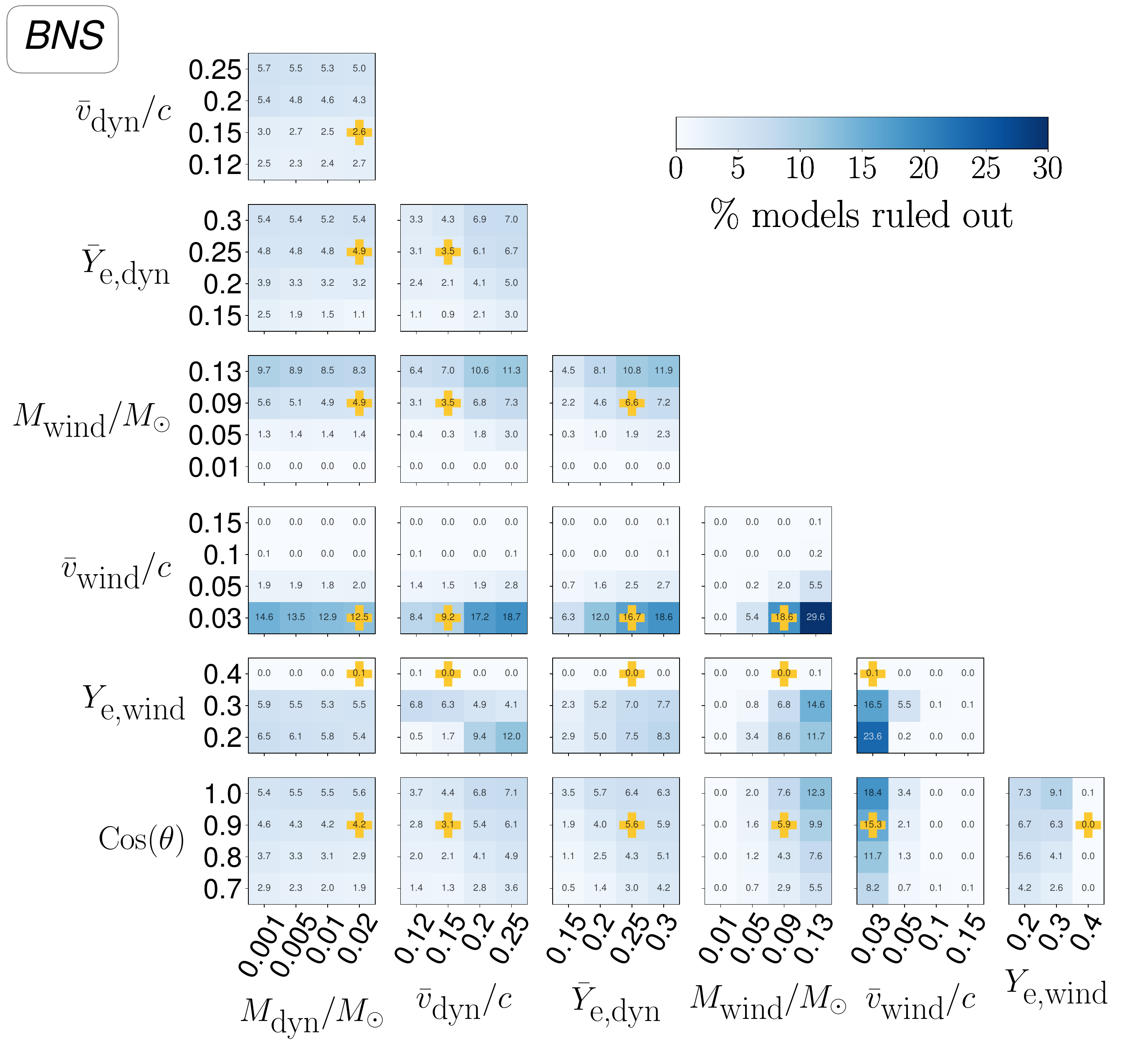}{0.48\textwidth}{(c) BNS model constrained by DECam observations}
          \fig{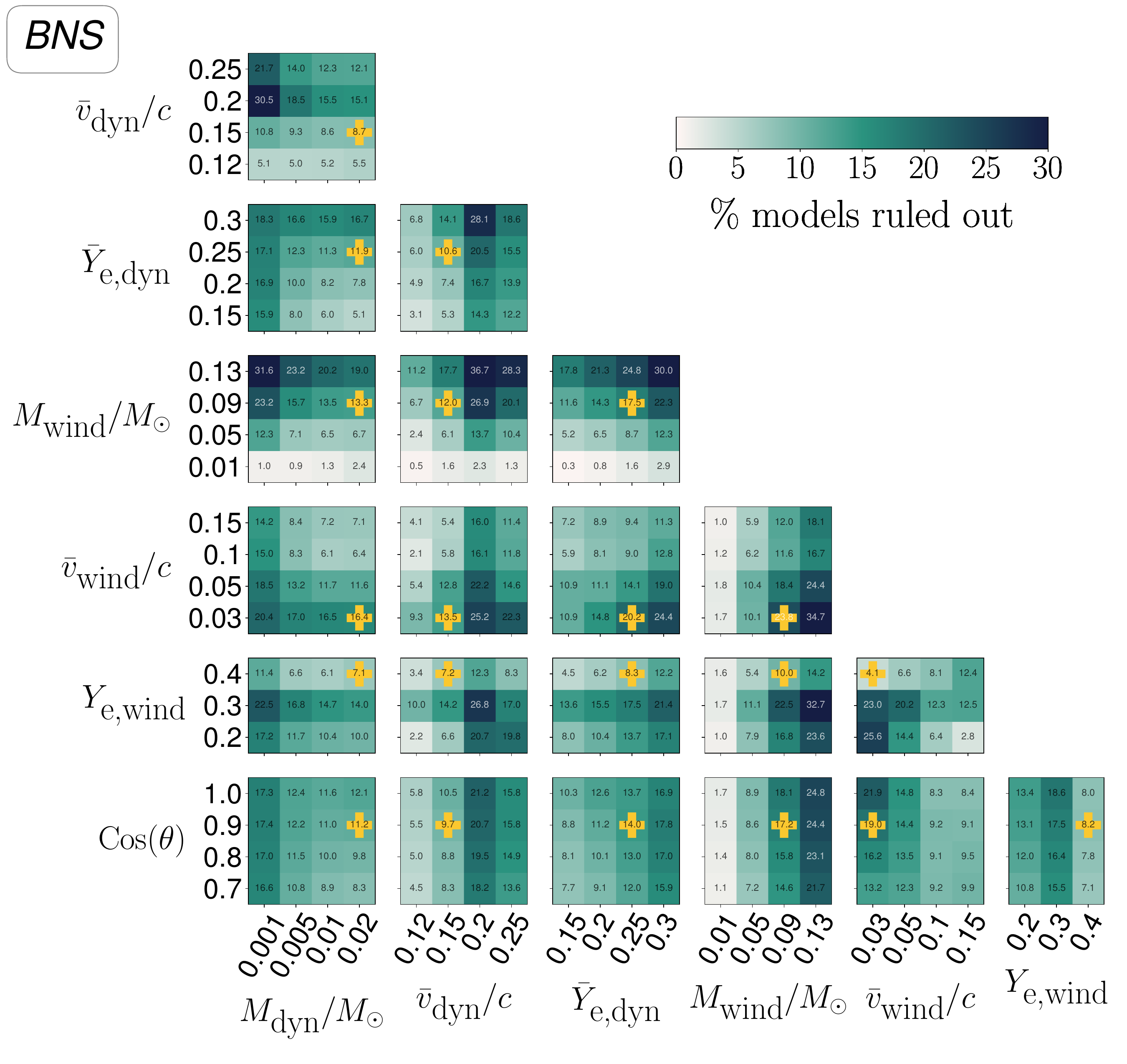}{0.48\textwidth}{(d) BNS model constrained by DECam and T80S observations}}

\caption{Corner plots similar to those in Fig. \ref{fig:model_constrain} showing the fraction of BHNS and BNS models ruled out by our observations, but here with a range of different viewing angles. For visualization purposes we only show models to viewing angles of $0-45$ deg, since from 45 deg and above the fractions stay roughly similar.}
\label{fig:model_constrain_with_angle}
\end{figure*}

In Figure \ref{fig:model_constrain} we show the kilonova model constraints for a fiducial viewing angle of 25 deg. In Figure~\ref{fig:model_constrain_with_angle} of this appendix, we show the same results but now considering a range of viewing angles between $0-45$ deg. Viewing angles of $45-90$ deg are roughly similar as they are dominated by faint emission from the lanthanide-rich dynamical ejecta and are less sensitive to the wind ejecta, which for edge-on systems is therefore mostly screened. In fact, it is clear from Figure~\ref{fig:model_constrain_with_angle} how the constraints of the wind ejecta parameters are significantly sensitive to variations in viewing angle, while the dynamical ejecta parameters are less sensitive.

\begin{deluxetable}{ccccc}
    \tablecaption{
        SALT spectroscopy for S250206dm.
        See the text for details on the instrumental configuration, which was identical for all spectra.
        \label{tab:salt}
    }
    \tablehead{
        TNS ID & SALT obs. date & Identified lines & Redshift & Notes
    }
    \startdata
        AT 2025bnp & 2025-02-14 & H$\alpha$, H$\beta$, O[III] & $0.1 \pm 0.01$ & Low SNR, increased uncertainty in line identification  \\
        AT 2025bnj & 2025-02-15 & H$\beta$, O[II], O[III] & $0.14 \pm 0.001$ & Observability conditions not met \\
        AT 2025bnj & 2025-02-16 & H$\beta$, O[II], O[III] & $0.14 \pm 0.001$ & Repeat observation \\
        AT 2025bnt & 2025-02-17 & - & - & No lines identified \\
        AT 2025bnh & 2025-02-22 & H$\beta$, O[II], O[III] & $0.14 \pm 0.001$ & - \\
    \enddata
\end{deluxetable}

\bibliography{S250206dm}{}
\bibliographystyle{aasjournal}



\end{document}